\documentclass[aip,pop,preprint,superscriptaddress]{revtex4-1}
\usepackage{setspace}
\usepackage{amsfonts,amsmath,amssymb}
\usepackage{txfonts}
\usepackage{graphics}
\usepackage{bm}
\usepackage{verbatim}
\usepackage{hyperref}
\usepackage{units}
\usepackage[T1]{fontenc}
\usepackage[latin9]{inputenc}
\usepackage{amsbsy}
\usepackage{mathrsfs}
\usepackage{esint}
 
\begin{document}

\title{The Hamiltonian structure and Euler-Poincar\'{e} formulation of the Vlasov-Maxwell and gyrokinetic
systems}

\author{J.~Squire}
\affiliation{Plasma Physics Laboratory, Princeton University, Princeton, New Jersey 08543, USA}
\author{H.~Qin}
\affiliation{Plasma Physics Laboratory, Princeton University, Princeton, New Jersey 08543, USA}
\affiliation{Dept.~of Modern Physics, University of Science and Technology of China, Hefei, Anhui 230026, China}
\author{W.~M.~Tang}
\affiliation{Plasma Physics Laboratory, Princeton University, Princeton, New Jersey 08543, USA}
\author{C. Chandre}
\affiliation{Centre de Physique Th\'eorique, CNRS -- Aix-Marseille Universit\'e, Campus de Luminy, 13009 Marseille, France}

\begin{abstract}
We present a new variational principle for the gyrokinetic system, similar to the Maxwell-Vlasov action 
presented in Ref.~\onlinecite{Cendra98themaxwell-vlasov}. The variational principle is in the Eulerian
frame and based on \emph{constrained} variations of the phase space fluid velocity and particle distribution 
function. Using a Legendre transform, we explicitly derive the field theoretic Hamiltonian structure 
of the system. This is carried out with a modified  Dirac theory of constraints, which is used to construct meaningful brackets from those obtained directly from Euler-Poincar\'{e} theory. Possible applications of these formulations include continuum geometric integration techniques, large-eddy simulation models and Casimir type stability methods.
\end{abstract}

\maketitle

\section{Introduction}

An inherent difficulty in studying the dynamics of magnetized plasmas
is the enormous separation of important time-scales present in many
physical systems of interest. Nonlinear gyrokinetic theory has become
an indispensable tool in these inquiries, as it removes the fastest
time-scales from the system, while keeping much of important physics
relevant to turbulent transport\cite{Brizard:2007p5792,Krommes:2012p8330,Lin:1998p8665}.
A particularly nice way to construct a gyrokinetic theory, pioneered
in Refs.~\onlinecite{Dubin:1983p9400,Littlejohn:1982p8584,hahm:1940},
is to use Lie-transforms to asymptotically change into co-ordinates
in which gyro-orbit dynamics are decoupled from the rest of the system.
A great advantage of this technique, aside from the entirely systematic
and formal procedure, is that the single particle equations are guaranteed
to be Hamiltonian, with associated conservation properties. Going
further, it is advantageous from both a philosophical and practical
standpoint to derive the entire system\emph{, }including both electromagnetic
fields and particles, from a single \emph{field-theoretic} variational
principle. These ideas were explored by Sugama\cite{Sugama:2000p9020}
and Brizard\cite{Brizard:2000p8804}, who derived gyrokinetic action
principles starting from Maxwell-Vlasov theories, as well as in previous
work in Refs.~\onlinecite{Similon:1985p9562,1987PhDT.......197B,Pfirsch:1985p9025,Pfirsch:1991p9026}. Some
advantages of this type of formulation are a much simplified derivation
of the gyrokinetic Maxwell's equations and exact energy-momentum conservation
laws through Noether's theorem. Field theories often admit many different
variational principles (e.g., for Maxwell-Vlasov see Refs.~\onlinecite{Pfirsch:1985p9025,Pfirsch:1991p9026,Ye:1992p8339,Brizard:2000p5815,Low:1958p5805,Fla:1994p9412,Cendra98themaxwell-vlasov}),
each with its own advantages and disadvantages. A good example is
the difference between Lagrangian and Eulerian actions; the former
being constructed in variables that follow particle motion and the
latter in variables at fixed points in phase space. It is interesting
to explore new types of variational principles, both for the general
understanding of the structure of the theory in question, and for practical
applications that may require an action of a particular form. 

In this work, we present a new gyrokinetic action principle in Eulerian
co-ordinates, using Euler-Poincar\'{e} reduction theory\cite{holm2009geometric,Holm:1998p9060,Newcomb1962}
on the Lagrangian action in Refs.~\onlinecite{Qin:2007p5801,Sugama:2000p9020}. In addition,
using the reduced Legendre transform and a modified version of the Dirac theory of constraints,
we derive field theoretic Poisson brackets, similar to the Vlasov-Maxwell\cite{Marsden:1982p9058,morrison:012104,ChandreBrackets2012,Morrison:1980:VMbracket,Morrison:1980p10549}
and Vlasov-Poisson\cite{1982AIPC...88...13M,ChandreBrackets2012} brackets. To our knowledge this is the
first explicit demonstration of the field theoretic Hamiltonian structure of the gyrokinetic
system (see Refs.~\onlinecite{morrison:012104,Morrison:2012p10456} for a different approach 
that has recently been used to write down a Poisson brackets for simplified gyrokinetic systems). 
Our derivation proceeds from the action principle in Ref.~\onlinecite{Sugama:2000p9020}
and its geometric formulation\cite{Qin:2007p5801}. We do not purport
to derive a gyrokinetic co-ordinate system, but rather formulate the
theory based on a given single particle Lagrangian. In this way, it
is trivial to extend  concepts to deal with more complex gyrokinetic
theories, for instance theories with self consistent, time-evolving
background fields\cite{Qin:2007p5801}. We then use the ideas in Ref.~\onlinecite{Cendra98themaxwell-vlasov}
to reduce the Lagrangian action to one in Eulerian co-ordinates, based
on symmetry under the particle-relabeling map from Lagrangian to Eulerian
variables. The variations for this new action principle in the Eulerian
frame are \emph{constrained, }and lead to the \emph{Euler-Poincar\'{e}
equations,} which are shown to give the standard gyrokinetic Vlasov
equation. In some ways the action principle is similar to that of
Brizard\cite{Brizard:2000p8804}, in that constrained variations must be used,
with both theories having a similar form for the variation of the
distribution function, $F$. Nevertheless, there are significant differences,
particularly that our principle is formulated in terms of the Eulerian phase
space fluid velocity and is in standard 6-D phase space, rather than
8-D extended phase space. The Eulerian gyrokinetic action of 
Refs.~\onlinecite{Pfirsch:2004p9371} is 
quite different to that presented here, with unconstrained variations in 
12-dimensional extended phase space and the use of Hamilton-Jacobi functions 
in the action functional. For more information on this approach to gyrokientic theory,
see Refs.~\onlinecite{CorreaRestrepo:2005p10302,CorreaRestrepo:2004p10301,CorreaRestrepo:2004p10303,CorreaRestrepo:2005p10307,Pfirsch:1985p9025,Pfirsch:1991p9026}.
Equipped with the Eulerian action, a reduced Legendre transform 
is performed\cite{Cendra98themaxwell-vlasov}, leading straightforwardly
to a Poisson bracket. However, this bracket must be \emph{reduced} to a constraint submanifold before
a meaningful form can be obtained, a process that is performed with a modified version of
the  Dirac theory of constraints\cite{Dirac:1950p9409,Chandre:2011p9324,Chandre:2010p9238}.
Finally, we show how to include the electromagnetic fields in the
bracket via second application of Dirac theory.

One of our primary motivations in this work is the possibility
of utilizing recent ideas from fluid mechanics to develop advanced
numerical tools for gyrokinetics. Of particular importance is the
idea of geometric integrators, which are designed to numerically conserve
various important geometrical properties of the physical system. For
instance, having a numerical algorithm that has Hamiltonian structure
can be very important, with profound consequences for the long-time
conservation properties\cite{Marsden:2001varint}. The theory
of finite dimensional geometric integrators is relatively well developed\cite{Marsden:2001varint}, including 
an application to single particle guiding center dynamics\cite{PhysRevLett.100.035006,li:052902,squire:052501}.
However, many aspects of the construction of field-theoretic geometrical integrators
are not as well understood, both for practical implementation and the deeper mathematical theory. 
One approach, which has yielded fruitful
results, is to discretize a variational principle and perform variations
on the discrete action to derive an integration scheme. Some examples of field theoretic
integrators constructed in this way are those for elastomechanics\cite{Lew2003_AVIs,Marsden:1998p5813},
electromagnetism\cite{Stern:2008p5799}, fluids and magnetohydrodynamics\cite{Pavlov:2009p5951,Gawlik:2011p5814},
and a particle-in-cell (PIC) scheme for the Vlasov-Maxwell system\cite{Squire:2012p9561}.
The results presented in this work would be used to construct a continuum Eulerian
gyrokinetic integrator, since our variational principle is in Eulerian
form. Analogously, a variational principle in Lagrangian form is used
to construct a Lagrangian (particle-in-cell) integrator\cite{Squire:2012p9561}.
We note that in discretizing a variational principle it is obviously not desirable 
to be in an extended phase space, unless these extra dimensions can somehow be 
removed after a discretization.  
As well as integrators, other potential applications of the formulation
presented here are the use in stability calculations with Casimir
invariants \cite{Holm:1985p8854,Kruskal:1958p10454} and the construction of regularized
models for large-eddy simulation\cite{Morel:2011p9388,PietarilaGraham:2011p9392,Bhat:2005p9397,Holm:2002p8805}.

It is significant to note at this point that as the power of modern supercomputing systems continues to advance at a rapid pace toward the exascale ($10^{18}$ floating point operations per second) and beyond, it is quite clear that the associated software development challenges are also increasingly formidable. On the emerging architectures memory and data motion will be serious bottlenecks as the required low-power consumption requirements lead to systems with significant restrictions on available memory and communications bandwidth. Consequently, it will be the case in multiple application domains that it will become necessary to re-visit key algorithms and solvers -- with the likelihood that new capabilities will be demanded in order to keep up with the dramatic architectural changes that accompany the impressive increases in compute power.  The key challenge here is to develop new methods to effectively utilize such dramatically increased parallel computing power.  Algorithms designed using geometric ideas could be very important 
as simulations of more complex systems are extended to longer times and enlarged spatial domains with high 
physics fidelity.

The article is organized as follows. In Section~\ref{sec:EP explanation}
we clarify the differences between Eulerian and Lagrangian action
principles for kinetic theories and explain the Euler-Poincar\'{e} formulation
of the Maxwell-Vlasov system\cite{Cendra98themaxwell-vlasov}. This
is done with as little reference to the formal mathematics as possible,
with the hope that readers unfamiliar with the concepts of Lie groups
and algebras will understand the general structure of the theory.
Section~\ref{sec:Gyrokinetic-variational-principle} explains the
construction of the gyrokinetic variational principle, starting from
a given single particle gyrokinetic Lagrangian. We give a brief derivation
of the Euler-Poincar\'{e} equations and show how these lead to a standard
form of the gyrokinetic equations. The Hamiltonian structure is dealt
with in Section~\ref{sec:The-Hamiltonian-formulation}. After formally
constructing a Poisson bracket from the Lagrangian, we describe how the modified
Dirac theory of constraints is used to reduce the
bracket to a meaningful form. Finally, numerical applications are briefly
discussed in Section~\ref{sec:Numerical-applications} and conclusions
given in Section~\ref{sec:Concluding-Remarks}. 

Throughout this article we use cgs units. In integrals and derivatives, $\bm{z}$ denotes all phase space 
variables, while $\bm{x}$ denotes just position space variables. Species labels are left out for
clarity and implied on the variables 
$F$ (or $f$), $m$, $e$, $\bm{U}$ and $\bm{M}$, respectively the distribution function, particle mass, particle charge,   
Eulerian fluid velocity and momenta conjugate to $\bm{U}$. Summation notation is utilized where applicable, 
with capital indices spanning $1\rightarrow6$ and lower case indices $1\rightarrow3$.

\section{Eulerian and Lagrangian kinetic variational principles\label{sec:EP explanation} }

When formulating a variational principle for a continuum fluid-type
theory, it is very important to specify whether Lagrangian or Eulerian
variables are being used. These notions can be confusing in kinetic
plasma theories, since one must consider the motion of the \emph{phase-space}
fluid. In addition, unlike the Euler fluid equations, the equations
of motion for kinetic plasma theories have the same form in Eulerian
and Lagrangian co-ordinates. Considering the Vlasov-Maxwell system
for simplicity, a Lagrangian description gives the equation of motion
at the position of a particle carried along by the flow (simply a
physical particle). One formulates a variational principle in terms
of the fields $\bm{x}\left(\bm{x}_{0},\bm{v}_{0},t\right)$, $\bm{v}\left(\bm{x}_{0},\bm{v}_{0},t\right)$,
which are the current position and velocity of an element of phase
space that was initially at $\left(\bm{x}_{0},\bm{v}_{0}\right)$.
The distribution function is of course just carried along by the Lagrangian
co-ordinates, i.e., $f\left(\bm{x}\left(\bm{x}_{0},\bm{v}_{0},t\right),\bm{v}\left(\bm{x}_{0},\bm{v}_{0},t\right)\right)=f_{0}\left(\bm{x}_{0},\bm{v}_{0}\right)$.
This type of formulation is the most natural for a kinetic theory,
since it is the logical continuum generalization of the action principle
for a collection of particles interacting with an electromagnetic
field. 

\begin{figure}

\begin{centering}
\includegraphics{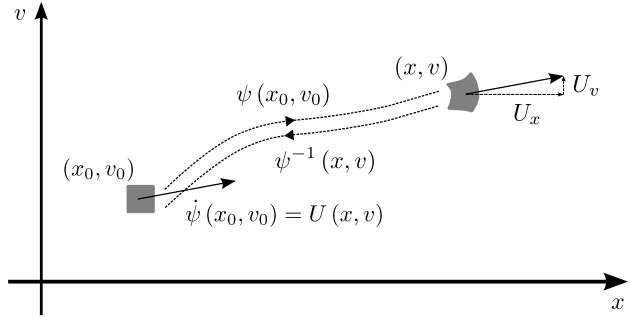}\caption{Illustration of the particle relabelling map, $\psi\left(\bm{x}_{0},\bm{v}_{0}\right)$
and its inverse for the one-dimensional Vlasov-Poisson system.\label{fig:psi map illustration}}

\par\end{centering}

\end{figure}

An Eulerian variational principle is formulated in terms of the velocity
of the phase space fluid at a fixed point, $\bm{U}$, without the
notion of where phase space density has been in the past. Thus, at
a point $\left(\bm{x},\bm{v}\right)$, the $\bm{x}$ component of
the fluid velocity is simply $\bm{v}$ (the co-ordinate),
while the $\bm{v}$ component is $\bm{E}+\bm{v}\times\bm{B}/c$. The
distribution function $f$, is \emph{advected }by $\bm{U}$, meaning
it is the solution to the differential equation $\partial_{t}f=-\mathcal{L}_{\bm{U}}f=-\bm{U}\cdot\nabla f$,
where $\bm{U}$ and $\nabla$ are in six-dimensional phase space. An illustration 
of these concepts is given in Figure~\ref{fig:psi map illustration} for the 1-D Poisson-Vlasov system 
(in 2-D phase space).  Finally,
we note that in discussing the distinction between Eulerian and Lagrangian
actions, we refer only to the plasma component of the variational
principle; the electromagnetic fields are always in Eulerian co-ordinates.%

\subsection{Euler-Poincar\'{e} reduction}\label{EP reduction explanation}

This section gives a very informal introduction to Euler-Poincar\'{e}
theory through a brief review of the Vlasov-Maxwell formulation presented
in Ref.~\onlinecite{Cendra98themaxwell-vlasov}. We purposefully do not
use the group theoretical notation of Refs.~\onlinecite{holm2009geometric,Holm:1998p9060,Cendra98themaxwell-vlasov} (e.g., the $\diamond$ and $\mathrm{ad}^{\star}$
operations) so as to introduce the general ideas to readers not familiar
with Lie groups and algebras. Euler-Poincar\'{e} type variational principles first 
appeared in Ref.~\onlinecite{Newcomb1962} in the context of magnetohydrodynamics.

The purpose of the Euler-Poincar\'{e} framework is to provide a straightforward
method to pass from a Lagrangian to an Eulerian action principle. The important
idea is that the Lagrangian is invariant under the right action of the particle relabeling transformation,
$\psi\left(\bm{x}_{0},\bm{v}_{0}\right)=\left(\bm{x}\left(\bm{x}_{0},\bm{v}_{0}\right),\bm{v}\left(\bm{x}_{0},\bm{v}_{0}\right)\right)$,
which maps plasma particles with initial position $\left(\bm{x}_{0},\bm{v}_{0}\right)$
to their current position $\left(\bm{x},\bm{v}\right)$. In essence, this invariance means that
we can eliminate the extraneous particle labeling information
and still have an equivalent system.
The map $\psi$ acts on $f$ on the right as $f=f_{0}\,\psi^{-1}$,
where $f_{0}$ is the initial distribution function. This equation
is simply $f\left(\bm{x}\left(\bm{x}_{0},\bm{v}_{0},t\right),\bm{v}\left(\bm{x}_{0},\bm{v}_{0},t\right)\right)=f_{0}\left(\bm{x}_{0},\bm{v}_{0}\right)$,
as discussed above. In the electromagnetic part of the Lagrangian $\psi$ does not act on the potentials
$\phi$ and $\bm{A}$, since electromagnetic dynamics 
must be independent of particle labeling. The phase space fluid
velocity in the Lagrangian frame is simply $\dot{\psi}\left(\bm{x}_{0},\bm{v}_{0}\right)$,
since this is the rate of change of $\left(\bm{x},\bm{v}\right)$
at the position $\left(\bm{x},\bm{v}\right)$. In contrast, the Eulerian
phase space fluid velocity is $\dot{\psi}\psi^{-1}$, since this operation
first takes $\left(\bm{x},\bm{v}\right)$ back to $\left(\bm{x}_{0},\bm{v}_{0}\right)$
with $\psi^{-1}$ and then gives the velocity at $\left(\bm{x},\bm{v}\right)$
with $\dot{\psi}\left(\bm{x}_{0},\bm{v}_{0}\right)$, see Figure~\ref{fig:psi map illustration}. 

The starting point for the reduction is a Lagrangian Lagrangian for
the Vlasov-Maxwell system. For instance, \begin{align}
L & =\sum_{s}\int d\bm{x}_{0}d\bm{v}_{0}\, f_{0}\left[\left(\frac{e}{c}\bm{A}\left(\bm{x}\right)+m\bm{v}\right)\cdot\dot{\bm{x}}-\frac{1}{2}m\bm{v}^{2}-e\phi\left(\bm{x}\right)\right]\nonumber \\
 & +\frac{1}{8\pi}\int d\bm{x}\left[\left|-\nabla\phi-\frac{\partial\bm{A}}{\partial t}\right|^{2}-\left|\nabla\times\bm{A}\right|^{2}\right],\label{eq:Full Low Lagrangian}\end{align}
which is very similar to the action principle of Low\cite{Low:1958p5805}. The
Vlasov equation follows from the standard Euler-Lagrange equations
for $\psi=\left(\bm{x},\bm{v}\right)$, \begin{equation}
\frac{d}{dt}\frac{\delta L}{\delta\dot{\psi}}-\frac{\delta L}{\delta\psi}=0,\label{eq:VM EL equations}\end{equation}
along with $f\left(\bm{x},\bm{v},t\right)=f_{0}\left(\bm{x}_{0},\bm{v}_{0}\right)$.
Maxwell's equations come from the Euler-Lagrange equations for $\bm{A}$
and $\phi$. This Lagrangian is invariant under the relabeling transformation $\psi$, i.e.,\begin{align}
L_{f_{0}}\left(\psi,\dot{\psi},\phi,\dot{\phi},\bm{A},\dot{\bm{A}}\right) & =L_{f_{0}\psi^{-1}}\left(\psi\psi^{-1},\dot{\psi}\psi^{-1},\phi,\dot{\phi},\bm{A},\dot{\bm{A}}\right)\nonumber \\
 & \equiv l\left(\bm{U},\phi,\dot{\phi},\bm{A},\dot{\bm{A}},F\right),\label{eq:Reduction of L}\end{align}
where $\bm{U}=\dot{\psi}\psi^{-1}$ is the Eulerian
fluid velocity, a vector field. In recognizing that the distribution
function is actually a phase space \emph{density}, we denote $F=f\, d\bm{x}\wedge d\bm{v}$.
Treating $F$ as 6-form rather than a scalar changes the form of certain
geometrical operations in the Euler-Poincar\'{e} equations {[}Eqs.~\eqref{eq:EP equations, f}
and \eqref{eq:EP equations, U}{]} and is very important for the gyrokinetic
Euler-Poincar\'{e} treatment (see Section~\ref{sec:Gyrokinetic-variational-principle}).
Practically speaking, to construct the reduced Lagrangian, $l,$ one simply
replaces $\left(\dot{\bm{x}},\dot{\bm{v}}\right)$ with $\bm{U}$, 
and considers $\bm{x}$ and $\bm{v}$
to be co-ordinates rather than fields. Thus,\begin{align}
l & =\sum_{s}\int F\left[\left(\frac{e}{c}\bm{A}+m\bm{v}\right)\cdot\bm{U}_{x}-\frac{1}{2}m\bm{v}^{2}-e\phi\left(\bm{x}\right)\right]\nonumber \\
 & +\frac{1}{8\pi}\int d\bm{x}\left[\left|-\nabla\phi-\frac{\partial\bm{A}}{\partial t}\right|^{2}-\left|\nabla\times\bm{A}\right|^{2}\right],\label{eq:Reduced Low Lagrangian}\end{align}
where $\bm{U}_{x}$ denotes the $\bm{x}$ components of $\bm{U}$.
The equations of motion are derived from the reduced Lagrangian $l$,
by considering how the unconstrained variations of $\psi$ (used to
derive the standard Euler-Lagrange equations) translate into \emph{constrained
}variations of $\bm{U}$ and $F$. This leads to variations of the
form 
\begin{equation}
\delta\bm{U}=\frac{\partial\eta}{\partial t}-\left[\bm{U},\eta\right],\;\; \delta F=-\mathcal{L}_{\eta}F,\label{eq:Reduced variations form}
\end{equation}
where $\eta$ is in the same space as $\bm{U}$
and vanishes at the endpoints; and $\left[\,,\right]$ is the standard
Lie bracket, $\bm{U}.\nabla\eta-\eta.\nabla\bm{U}$. Evolution of $F$ is 
given by the advection equation 
\begin{equation}
\frac{\partial F}{\partial t}+ \mathcal{L}_{\bm{U}}F=0 \label{eq:EP equations, f},
\end{equation}
which arises from the equation $f\left(\bm{x},\bm{v},t\right)=f_0\left(\bm{x}_0,\bm{v}_0\right)$.
Variation of
$\int dt\, l$ with $\delta\bm{U}$ and $\delta F$ leads to the \emph{Euler-Poincar\'{e}
equations},
\begin{equation}
\frac{\partial}{\partial t}\frac{\delta l}{\delta\bm{U}}  =-\mathcal{L}_{\bm{U}}\frac{\delta l}{\delta\bm{U}}+F\nabla\frac{\delta l}{\delta F}\label{eq:EP equations, U},\end{equation}
 where $\delta l/\delta\bm{U}$ is
a 1-form density. We give straightforward derivations of Eqs.~\eqref{eq:Reduced variations form} 
and \eqref{eq:EP equations, U}
in Section~\ref{sub:EP-Gyrokinetic-variational} below. Since $F$
is a 6-form, $\mathcal{L}_{\bm{U}}F=\nabla\cdot\left(F\bm{U}\right)$
and Eq.~\eqref{eq:EP equations, f} is the conservative form of the
Vlasov equation (see Section~\ref{sub:EP-Gyrokinetic-variational}
for more information). The equations for $\bm{A}$ and $\phi$ are
just the standard Euler-Lagrange equations. Calculation of Eq.~\eqref{eq:EP equations, U}
with the Vlasov-Maxwell reduced Lagrangian {[}Eq.~\eqref{eq:Reduced Low Lagrangian}{]}
leads to \begin{equation}
\bm{U}_{x}=\bm{v},\:\bm{U}_{v}=\bm{E}+\frac{1}{c}\bm{v}\times\bm{B},\label{eq:EP solution, VM}\end{equation}
as expected. The fact that there is no need to solve differential
equations for components of $\bm{U}$ is related to the strong degeneracy
in the system (see Sections~\ref{sec:Gyrokinetic-variational-principle}
and \ref{sec:The-Hamiltonian-formulation} below). 

To obtain the Hamiltonian or \emph{Lie-Poisson} form of the equations,
one performs a reduced Legendre transform as,
\begin{equation}
h=\left\langle \bm{M},\bm{U}\right\rangle+\int d\bm{x}\, \bm{A}\cdot \frac{\delta l}{\delta \dot{\bm{A}}} -l,
\label{eq:legendre trans defn}\end{equation}
where $\bm{M}=\delta l/\delta\bm{U}$ and the inner product $\left\langle \, , \right\rangle$ is  
integration over phase space [see Eq.~\eqref{eq:Full EP deivation}]. (A thorough treatment of the 
degeneracies of the system is given in Ref. \onlinecite{Cendra98themaxwell-vlasov}.) 
It is then straightforward to show that
\begin{align}
\left\{ \Gamma,\Theta\right\} _{LP} & = -\sum_s \int d\bm{z}\bm{M}\cdot\left[\frac{\delta\Gamma}{\delta\bm{M}},\frac{\delta\Theta}{\delta\bm{M}}\right]\nonumber \\
+ & \sum_s \int d\bm{z}F\left(\frac{\delta\Theta}{\delta\bm{M}}\cdot\nabla\frac{\delta\Gamma}{\delta F}-\frac{\delta\Gamma}{\delta\bm{M}}\cdot \nabla \frac{\delta\Theta}{\delta F}\right)\nonumber \\
- &\,4\pi c \int d\bm{x}\left(\frac{\delta\Gamma}{\delta\bm{A}}\cdot\frac{\delta\Theta}{\delta\bm{E}}-\frac{\delta\Theta}{\delta\bm{A}}\cdot\frac{\delta\Gamma}{\delta\bm{E}}\right),\label{eq:LP bracket definition}
\end{align}
is an infinite dimensional Poisson bracket for the system; that is, $\dot{\bm{M}}=\left\{ \bm{M},h\right\} $, 
$\dot{F}=\left\{ F,h\right\} $, $\dot{\bm{E}}=\left\{ \bm{E},h\right\} $ and  $\dot{\bm{A}}=\left\{ \bm{A},h\right\} $ are formally
the same as the Euler-Poincar\'{e} equations (using the generalized Legendre transform 
of Ref.~\onlinecite{Cendra98themaxwell-vlasov}), and the Jacobi identity
is satisfied. Nevertheless, this manifestation of the bracket has major problems.
In particular, the meaning of functional derivatives with respect to the  
$\bm{M}$ variables can be unclear, since these are constrained due to the linearity of the 
Lagrangian in $\bm{U}$.
To overcome these problems and
formulate a meaningful bracket on the space of plasma densities and electromagnetic fields, 
we use a modified version of the Dirac theory of constraints\cite{Dirac:1950p9409,Pfirsch:1991p9026,Morrison20091747}. For the convenience of the reader a very brief 
overview of standard Dirac theory in Appx.~\ref{Appendix DC}, while the modified version
used in parts of this work is covered in Appx.~\ref{Appendix ModDC}.

The relevant constraints are\begin{align}
\Phi_{i}= & M_{i}-F\left(\frac{e}{c}A_{i}+mv_{i}\right)=0,\quad i=1\rightarrow3,\nonumber \\
 & \Phi_{j}=M_{j}=0,\quad j=4\rightarrow6.\label{eq:VM dirac constraints}\end{align}
We
form the constraint matrix $\mathcal{C}_{IJ}\left(\bm{z},\bm{z}'\right)=\left\{ \Phi_{I}\left(\bm{z}\right),\Phi_{J}\left(\bm{z}'\right)\right\} $
and construct the inverse according to the procedure detailed in Appxs.~\ref{Appendix DC} and \ref{Appendix ModDC}. This comes out to be,
\begin{align}
\mathcal{C}_{IJ}^{-1} & \left(\bm{z},\bm{z}'\right)=\frac{1}{mF\left(\bm{z}\right)}\delta\left(\bm{z}-\bm{z}'\right)\delta_{ss'}\nonumber \\
\times & \left(\begin{array}{cccccc}
0 & 0 & 0 & 1 & 0 & 0\\
0 & 0 & 0 & 0 & 1 & 0\\
0 & 0 & 0 & 0 & 0 & 1\\
-1 & 0 & 0 & 0 & \frac{e}{mc}B_{z} & -\frac{e}{mc}B_{y}\\
0 & -1 & 0 & -\frac{e}{mc}B_{z} & 0 & \frac{e}{mc}B_{x}\\
0 & 0 & -1 & \frac{e}{mc}B_{y} & -\frac{e}{mc}B_{x} & 0\end{array}\right),\label{eq:VM Cinv mat}
\end{align}
which is simply the single particle Poisson matrix (multiplied by $\delta \left(\bm{z}-\bm{z}'\right)/F$) and $\bm{B}$ 
is a function of $\bm{z}$.
We will see a similar connection to the single particle Poisson bracket in the reduction of the gyrokinetic bracket.
We then use Eq.~\eqref{eq:Dirac Brak Defn} and restrict the functionals $\Gamma$ and $\Theta$
to not depend $\bm{M}$, i.e., $\delta\Gamma/\delta\bm{M}=0$ (see Appx.~\ref{Appendix ModDC}). 

The final result, including a change of variables from $\bm{A}$ to $\bm{B}$, is the
Poisson bracket for the Mawell-Vlasov system,\begin{align}
\left\{ \Gamma,\Theta\right\}  & =\sum_{s}\frac{1}{m}\int d\bm{z}F\left(\frac{\partial\Gamma_{F}}{\partial\bm{x}}\cdot\frac{\partial\Theta_{F}}{\partial\bm{v}}-\frac{\partial\Theta_{F}}{\partial\bm{x}}\cdot\frac{\partial\Gamma_{F}}{\partial\bm{v}}\right)\nonumber \\
 & +\sum_{s}\frac{e}{cm^{2}}\int d\bm{z}F\bm{B}\cdot\frac{\partial\Gamma_{F}}{\partial\bm{v}}\times\frac{\partial\Theta_{F}}{\partial\bm{v}}\nonumber \\
 & +4\pi\sum_{s}\frac{e}{m}\int d\bm{z}\left[\Theta_{F}\frac{\partial}{\partial\bm{v}}\cdot\left(F\frac{\delta\Gamma}{\delta\bm{E}}\right)-\Gamma_{F}\frac{\partial}{\partial\bm{v}}\cdot\left(F\frac{\delta\Theta}{\delta\bm{E}}\right)\right]\nonumber \\
 & +4\pi c\int d\bm{x}\left(\frac{\partial\Gamma}{\partial\bm{E}}\cdot\nabla\times\frac{\delta\Theta}{\delta\bm{B}}-\frac{\partial\Theta}{\partial\bm{E}}\cdot\nabla\times\frac{\delta\Gamma}{\delta\bm{B}}\right).\label{eq:Full VM bracket}\end{align}
This bracket is identical to that published previously, initially proposed in Ref.~\onlinecite{Morrison:1980:VMbracket} with a correction given later in Ref.~\onlinecite{Marsden:1982p9058}.
The derivation above explicitly shows the link between this and the work of 
Ref.~\onlinecite{Cendra98themaxwell-vlasov}. 
There is a slight taint on the validity of this bracket in that it requires $\nabla\cdot\bm{B}=0$
for the Jacobi identity to be satisfied (as shown in Refs.~\onlinecite{1982AIPC...88...13M,morrison:012104} through
direct calculation). Recently, this obstruction has been partially 
fixed using the Dirac theory of constraints\cite{ChandreBrackets2012}. 
Somewhat more detail is given for the derivation of the gyrokinetic 
bracket  (see Section \ref{sec:The-Hamiltonian-formulation}), which proceeds in a very similar 
manner.

Euler-Poincar\'{e} reduction is perhaps more natural when applied to fluid
systems\cite{Newcomb1962, Holm:1998p9060}. In this case, there are fewer degeneracies and the Lagrangian/Eulerian distinction 
is more obviously relevant (e.g.~one does not \emph{measure} phase space fluid velocities, in 
contrast to the Eulerian fluid velocity of a fluid system). 
Recently, similar ideas have been applied to general reduced fluid
and hybrid models in plasma physics\cite{Holm2012_EPplasma,Brizard2055_Noether,0741-3335-55-3-035001}.

\section{Gyrokinetic variational principle\label{sec:Gyrokinetic-variational-principle}}

Our starting point is the geometric approach to gyrokinetic theory
advocated in Ref.~\onlinecite{Qin:2007p5801}. The general idea is to construct
a field theory, including electromagnetic potentials, from the particle
Poincar\'{e}-Cartan 1-form, $\gamma$. This approach is conceptually very
simple; once the interaction of quasi-particles with the electromagnetic
field is specified, particle and field equations follow in straightforward
and transparent way via the Euler-Lagrange equations. With any desired
approximation (e.g., expansion in gyroradius), energetically self-consistent
equations are easily obtained without necessitating the use of the
pullback operator. The use of these ideas in gyrokinetic simulation
has been advocated in, for instance Refs.~\onlinecite{Scott:2010p9381,Scott:2010p9978}.
In this article we consider the particle 1-form $\gamma$ as given,
its derivation can be found in Refs.~\onlinecite{Sugama:2000p9020,Brizard:2007p5792,Qin:2007p5801} 
among other works. 

The Poincar\'{e}-Cartan form $\gamma$ in 7-D phase space, $P$, (including
time) defines particle motion through Hamilton's equation, \begin{equation}
i_{\tau}d\gamma=0,\label{eq:SP hamiltons equations}\end{equation}
which is derived from stationarity of the action $\mathcal{A}_{sp}=\int\gamma$.
Here $\tau$ is a vector field whose integrals define particle trajectories
(including the time component) and $i$ denotes the inner product.
Note that $\gamma$ is essentially just $L_{sp}dt$, where $L_{sp}$
is the standard Lagrangian; that is, for $\gamma=\gamma_{\alpha}dz^{\alpha}-Hdt$,
the Lagrangian is simply $L_{sp}=\gamma_{\alpha}\dot{z}^{\alpha}-H$.
To construct a field theory, $\gamma$ is used to define the Liouville
6-form,\begin{equation}
\Omega_{T}=-\frac{1}{3!}d\gamma\wedge d\gamma\wedge d\gamma.\label{eq:Louiville 6-form defn}\end{equation}
The Liouville theorem of phase space volume conservation is then simply,
$\mathcal{L}_{\tau}\Omega_{T}=0$. Introducing the distribution function
of particles in phase space $f$, the field theory action for the
interaction of a field of particles with the electromagnetic field
is \begin{equation}
\mathcal{A}=4\pi\int f\Omega_{T}\wedge\gamma+\int dx\mathcal{L}_{EM},\label{eq:Geometric Lagranian action}\end{equation}
where $\mathcal{L_{EM}}$ is the electromagnetic Lagrangian density.
In this action $\gamma$ is in the Lagrangian frame. For example,
in Cartesian position and velocity space, $f\Omega$ is in $\left(\bm{x}_{0},\bm{v}_{0}\right)$
co-ordinates, while $\gamma$ is in $\left(\bm{x}\left(\bm{x}_{0},\bm{v}_{0}\right),\bm{v}\left(\bm{x}_{0},\bm{v}_{0}\right)\right)$
co-ordinates. 

Unless general relativity is important,
one can choose $\tau$ to be of the form $\tau=\partial/\partial t+\tau_{Z}$\footnote{If space-time
is curved, the $\partial/\partial t$ component of $\tau$ could have dependence on 
the phase-space co-ordinates. When this is true, one can not work in 6-D phase space with a separate
treatment of time},
where $\tau_{Z}$ has no time component, and consider 6-D phase space.
Defining $\Omega$ to be the $d\bm{X}\wedge d\bm{P}$ component of
$\Omega_{T}$ (i.e., no $\wedge dt$), the $d\bm{X}\wedge d\bm{P}$
component of $\mathcal{L}_{\tau}\Omega_{T}$ is the standard Liouville
theorem of phase space volume conservation,\begin{equation}
\frac{\partial}{\partial t}\Omega+\mathcal{L}_{\tau_{Z}}\Omega=0.\label{eq:Standard Louiville theorem}\end{equation}
We can simplify the variational principle by considering $\gamma$
to be $L_{sp}dt$ and carrying out the wedge product. This type of
procedure provides a generalization of the original variational principle
of Low\cite{Low:1958p5805} [e.g., Eq.~\eqref{eq:Full Low Lagrangian}]
to arbitrary particle-field interaction. 

We now specialize to a general gyrokinetic form for the particle Lagrangian,
\begin{equation}
\gamma=\frac{e}{c}\bm{A}^{\dagger}\left(\bm{X}\right)\cdot d\bm{X}+\frac{mc}{e}\mu d\theta-Hdt,\label{eq:GK SP gamma}\end{equation}
with \begin{equation}
\bm{A}^{\dagger}\left(\bm{X}\right)=\bm{A}+\frac{mc}{e}u\bm{b}-\frac{mc^{2}}{e^{2}}\mu\left(\bm{R}+\frac{1}{2}\bm{b}\,\bm{b}\cdot\nabla\times\bm{b}\right).\label{eq:Full Adag defn}\end{equation}
Here, $\bm{X}$ is the gyrocenter position, $u$ the gyrocenter parallel
velocity co-ordinate, $\mu$ the conserved magnetic moment and $\theta$
the gyrophase.  The
vector field $\bm{A}\left(\bm{X}\right)$ is the vector potential
of the background magnetic field and $\bm{b}\left(\bm{X}\right)$
is the background magnetic field unit vector. These fields will not
be considered variables in the field theory action. The vector $\bm{R}\left(\bm{X}\right)=\nabla\bm{e}_{1}\cdot\bm{e}_{2}$,
where $\bm{e}_{1}\left(\bm{X}\right)\perp\bm{e}_{2}\left(\bm{X}\right)\perp\bm{b}\left(\bm{X}\right)$,
is necessary for gyrogauge invariance of the Lagrangian, i.e., invariance
with respect to a change in the definition of the $\theta$ co-ordinate.
Eq.~\eqref{eq:GK SP gamma} is accurate to first order in $\epsilon_{B}$,
the ratio of the gyroradius to the scale length of the magnetic field\cite{Brizard:2007p5792}.
The single particle Hamiltonian, $H=\frac{1}{2}mu^{2}+\mu B\left(\bm{X}\right)+H_{gy}$,
contains both the guiding center contribution, $\frac{1}{2}mu^{2}+\mu B\left(\bm{X}\right)$,
and the gyrocenter contribution from the fluctuating fields, $H_{gy}$.
For most of this article $H_{gy}$ will be taken to be a general function
of $\left(\bm{X},u,\mu\right)$. Different forms exist in the literature,
depending on desired accuracy and fluctuation model used. For instance,
in Ref.~\onlinecite{Sugama:2000p9020}, $H_{gy}$ is given to second order
in $\epsilon_{\delta}$ (the ratio of the magnitudes of the fluctuating
fields, $\phi_{1}$ and $\bm{A}_{1}$, to the background field) as\begin{align}
H_{gy} & =e\left\langle \psi\left(\bm{X}+\bm{\rho}\right)\right\rangle +\frac{e^{2}}{2mc^{2}}\left\langle \left|\bm{A}\left(\bm{X}+\bm{\rho}\right)\right|^{2}\right\rangle \label{eq:Full Hgy from Sugama}\\
 & -\frac{e}{2}\left\langle \left\{ \tilde{S}_{1},\tilde{\psi}\right\} \right\rangle ,\nonumber \end{align}
where $\psi=\phi_{1}-\frac{1}{c}\bm{v}\cdot\bm{A}_{1}$, $\bm{\rho}$
is difference between the particle and gyrocenter positions, and $\left\langle \right\rangle $
denotes an average over $\theta$. The tilde in $\left\langle \left\{ \tilde{S}_{1},\tilde{\psi}\right\} \right\rangle $
denotes the gyrophase dependent part of a function and $S_{1}$ is
a gauge function associated with the first order gyrocenter perturbation,
$\left\langle \psi\left(\bm{X}+\bm{\rho}\right)\right\rangle $. Eq.~\eqref{eq:GK SP gamma}
is the standard gyrokinetic single particle Lagrangian in \emph{Hamiltonian
}form\cite{Brizard:2007p5792}, meaning all the fluctuating field
perturbations are in the Hamiltonian part ($dt$ component) of $\gamma$.
This is the form most suitable for computer simulation\cite{Brizard:2007p5792,Scott:2010p9381}
and also has the advantage of having the same Poisson structure as the guiding center equations.

We now construct a field theory from the single particle Lagrangian Eq.~\eqref{eq:GK SP gamma}, 
a process often referred to as \emph{lifting}\cite{morrison:012104,Chandre:2012p10404,Morrison:2012p10456}.
We first calculate the phase space
component ($d\bm{X}\wedge du\wedge d\mu\wedge d\theta$ component)
of the volume element $\Omega=-\frac{1}{3}d\gamma\wedge d\gamma\wedge d\gamma$.
This is simply $B_{\parallel}^{\dagger}/m=\bm{b}\cdot\bm{B}^{\dagger}/m$,
with $\bm{B}^{\dagger}=\nabla\times\bm{A}^{\dagger}$, i.e., the standard
guiding center Jacobian. In co-ordinates, the variational principle
Eq.~\eqref{eq:Geometric Lagranian action} is then simply,\begin{align}
\mathcal{A} & =\int dt\, L_{GK}\nonumber \\
 & =\sum_{s}\int dt\int d\bm{X}_{0}\wedge du_{0}\wedge d\mu_{0}\wedge d\theta_{0}\frac{1}{m}B_{\parallel}^{\dagger}f_{0}\nonumber \\
 & \times\left[\frac{e}{c}\bm{A}^{\dagger}\cdot\dot{\bm{X}}+\frac{mc}{e}\mu\dot{\theta}-H\right]+\int dt\, L_{EM},\label{eq:Lagrangian GK action}\end{align}
which is essentially the original gyrokinetic variational principle
of Sugama\cite{Sugama:2000p9020}. $L_{EM}$ should be chosen as \begin{equation}
L_{EM}=\frac{1}{8\pi}\int d\bm{x}\left(\left|\nabla\phi_{1}\right|^{2}-\left|\nabla\times\left(\bm{A}+\bm{A}_{1}\right)\right|^{2}\right),\label{eq:EM Lagrangian part}\end{equation}
where $\bm{x}=\bm{X}+\bm{\rho}$,  so 
the Amp\`{e}re-Poisson system is obtained, removing fast time-scale electromagnetic
waves.

\subsection{Eulerian Gyrokinetic variational principle\label{sub:EP-Gyrokinetic-variational}}

We proceed in the reduction of the gyrokinetic variational principle,
Eq.~\eqref{eq:Lagrangian GK action}, in a very similar way to the
Vlasov-Maxwell case (Section \ref{sec:EP explanation}). The advected
parameter is the 6-form $f\Omega=d\bm{X}\wedge du\wedge d\mu\wedge d\theta\, B_{\parallel}^{\dagger}f/m \equiv\hat{F}d\bm{X}\wedge du\wedge d\mu\wedge d\theta$.
$\hat{F}$ is often considered a function for simplicity of notation, but
it is understood that operations should be carried out as for a 6-form,
e.g., $\mathcal{L}_{\bm{U}}\hat{F}=\nabla\cdot\left(\hat{F}\bm{U}\right)$
rather than $\mathcal{L}_{\bm{U}}\hat{F}=\bm{U}\cdot\nabla\hat{F}$.
The connection to the standard distribution function is provided by
the Liouville theorem; the advection equation \begin{equation}
\partial_{t}\left(f\Omega\right)+\mathcal{L}_{\bm{U}}\left(f\Omega\right)=0,\label{eq:GK 6-form advection}\end{equation}
coupled with Liouville's theorem, Eq.~\eqref{eq:Standard Louiville theorem},
implies \begin{equation}
\partial_{t}f+\mathcal{L}_{\bm{U}}f=0,\label{eq:Advection of f scalar}\end{equation}
which is the standard Vlasov equation. 

Operating on Eq.~\eqref{eq:Lagrangian GK action} on the right with the particle-relabeling map, $\psi^{-1}$, leads to the reduced Lagrangian 
\begin{align}
l_{GK} & \left(\bm{U},\phi_{1},\bm{A}_{1},\hat{F}\right)=L_{GK}\left(\psi\psi^{-1},\dot{\psi}\psi^{-1},\phi_{1},\bm{A}_{1},f_{0}\Omega\psi^{-1}\right)\nonumber \\
  =&\sum_{s}\int d\bm{X}dud\mu d\theta\,\hat{F}\left(\frac{e}{c}\bm{A}^{\dagger}\cdot\bm{U}_{X}+\frac{mc}{e}\mu U_{\theta}-H\right)\nonumber \\
 & +\frac{1}{8\pi}\int d\bm{x}\left(\left|\nabla\phi_{1}\right|^{2}-\left|\nabla\times\left(\bm{A}+\bm{A}_{1}\right)\right|^{2}\right),\label{eq:GK EP lagrangian}
 \end{align}
where $\bm{U}_{X}$ and $U_{\theta}$ are the $\bm{X}$ and $\theta$
components of the Eulerian fluid velocity $\bm{U}$. 

The unconstrained variations in the Lagrangian frame, $\delta\psi$ lead to constrained variations in the Eulerian frame by defining\cite{Newcomb1962, Holm:1998p9060}
\begin{equation}
\bm{\eta}\left(\bm{z},t\right)=\delta\psi\left(\bm{z}_{0},t\right),
\end{equation}
or equivalently $\bm{\eta}=\delta\psi\psi^{-1}$. Recalling $\bm{U}\left(\bm{z},t\right)=\dot{\psi}\left(\bm{z}_{0},t\right)$, one then calculates $d\bm{\eta}/dt$ and $\delta\bm{U}$, giving
\begin{align}
\delta\dot{\psi}\left(\bm{z}_{0},t\right)&=\frac{\partial\bm{\eta}\left(\bm{z},t\right)}{\partial t}+\dot{z}^{j}\frac{\partial\bm{\eta}\left(\bm{z},t\right)}{\partial z^{j}},\\\delta\dot{\psi}\left(\bm{z}_{0},t\right)&=\delta\bm{U}\left(\bm{z},t\right)+\delta z^{j}\frac{\partial\bm{U}\left(\bm{z},t\right)}{\partial z^{j}},
\label{eq:delta psi dot variations}
\end{align}
which is solved for 
\begin{equation}
\delta\bm{U}=\frac{\partial\bm{\eta}}{\partial t}+\bm{U}\cdot\nabla_{z}\bm{\eta}-\bm{\eta}\cdot\nabla_{z}\bm{U},
\label{eq:U variations derivation}
\end{equation}
giving the variational form stated in Eq.~\eqref{eq:Reduced variations form}. Similarly, using $f\Omega\left(\bm{z}\right)=f_{0}\Omega_{0}\left(\bm{z}_{0}\right)$ and $\delta\left(f_{0}\Omega_{0}\right)=0$, one obtains 
\begin{equation}
\delta\left(f\Omega\right)=-\mathcal{L}_{\eta}\left(f\Omega\right).
\label{eq:F variations derivation}
\end{equation}
Using Eqs.~\eqref{eq:U variations derivation} and \eqref{eq:F variations derivation}, we give a basic derivation of the Euler-Poincar\'{e} equations for a general
Lagrangian with an advected volume form, $\hat{F}$.\begin{align}
\delta\int dt & \, l=\int dt\left[\left\langle \frac{\delta l}{\delta\bm{U}},\delta\bm{U}\right\rangle _{\mathfrak{g}}+\left\langle \frac{\delta l}{\delta\hat{F}},\delta\hat{F}\right\rangle _{V}\right]\nonumber \\
= & \int dt\left[\left\langle \frac{\delta l}{\delta\bm{U}},\left(\frac{\partial\eta}{\partial t}+\left[\bm{U},\eta\right]\right)\right\rangle _{\mathfrak{g}}+\left\langle \frac{\delta l}{\delta\hat{F}},\mathcal{L}_{\eta}\hat{F}\right\rangle _{V}\right]\nonumber \\
= & \int dt\sum_{s}\int d\bm{X}dud\mu d\theta\left\{ -\frac{\partial}{\partial t}\frac{\delta l}{\delta\bm{U}}\eta^{i}-\left[\frac{\delta l}{\delta U^{j}}\partial_{i}U^{j}\right.\right.\nonumber \\
 & \left.\left.+\partial_{j}\left(\frac{\delta l}{\delta U^{i}}U^{j}\right)\right]\eta^{i}+\eta^{i}\hat{F}\partial_{i}\frac{\delta l}{\delta\hat{F}}\right\} \nonumber \\
= & \int dt\left\langle -\frac{\partial}{\partial t}\frac{\delta l}{\delta\bm{U}}-\mathcal{L}_{\bm{U}}\frac{\delta l}{\delta\bm{U}}+\hat{F}\nabla\frac{\delta l}{\delta\hat{F}},\,\eta\right\rangle _{\mathfrak{g}},\label{eq:Full EP deivation}\end{align}
giving Eq.~\eqref{eq:EP equations, U} since $\eta$ is arbitrary.
The two brackets used are defined as $\left\langle \mu,\xi\right\rangle _{\mathfrak{g}}=\sum_{s}\int d\bm{X}dud\mu d\theta\,\mu_{i}\xi^{i}$
between a 1-form density $\mu d\bm{z}$ and a vector field $\xi$;
and $\left\langle f,\hat{F}\right\rangle _{v}=\sum_s \int f\hat{F}$ between
a function $f$ and a volume form $\hat{F}$. Integration by parts
is used, with boundary terms dropped, in arriving at the third line.
Note that $\hat{F}$ sometimes includes the volume element (lines
1 and 2) and sometimes does not (lines 3 and 4). A more precise derivation
can be found in Refs.~\onlinecite{Cendra98themaxwell-vlasov,holm2009geometric}.

It is now simple to write down the equations of motion for $\bm{U}$,
using\begin{align}
 & \frac{\delta l_{GK}}{\delta\bm{U}_{X}}=\frac{e}{c}\hat{F}\bm{A}^{\dagger},\:\frac{\delta l_{GK}}{\delta U_{\theta}}=\frac{mc}{e}\mu\hat{F}\nonumber \\
 & \frac{\delta l_{GK}}{\delta U_{u}}=\frac{\delta l_{GK}}{\delta U_{\mu}}=0.\label{eq:lGK functional derivatives}\end{align}
The derivation is carried out without assumptions about the form of
$\bm{U}$ (e.g., lack of $\theta$ dependence) and the $\hat{F}$
advection equation {[}Eq.~\eqref{eq:GK 6-form advection}{]} is used
to cancel time derivatives. We illustrate the general form of the calculation with the $\delta l/\delta U_{X}^{i}$ component of Eq.~\eqref{eq:EP equations, U},
\begin{align}
\frac{\partial}{\partial t} & \left(\frac{e}{c}\hat{F}A_{i}^{\dagger}\right)=-\frac{e}{c}A_{i}^{\dagger}\frac{\partial}{\partial Z^{J}}\left(\hat{F}U^{J}\right)-\frac{e}{c}\hat{F}U^{J}\frac{\partial A_{i}^{\dagger}}{\partial Z^{J}}\nonumber \\
- & \frac{e}{c}\hat{F}A_{j}^{\dagger}\frac{\partial U^{j}}{\partial X^{i}}-\frac{mc}{e}\mu\hat{F}\frac{\partial U_{\theta}}{\partial X^{i}}+\hat{F}\frac{\partial}{\partial X^{i}}\left(\frac{e}{c}\bm{A}^{\dagger}\cdot\bm{U}_{X}\right.\nonumber \\
+ & \left.\frac{mc}{e}\mu U_{\theta}-\frac{1}{2}mu^{2}-\mu B-H_{gy}\right).\label{eq:Full EP equation for X cmpts}\end{align}
The first two terms in Eq.~\eqref{eq:Full EP equation for X cmpts}
add to zero due to the advection equation, while the terms involving
$U_{\theta}$ cancel. Rearranging and expanding the divergence term
leads to \begin{equation}
\frac{e}{c}\bm{U}_{X}\times\bm{B}^{\dagger}-mU_{u}\bm{b}-\mu\nabla B-\nabla H_{gy}=0,\label{eq:UX equation before dotting}\end{equation}
which gives\begin{equation}
U_{u}=-\frac{\bm{B}^{\dagger}}{mB_{\parallel}^{\dagger}}\cdot\left(\mu\nabla B+\nabla H_{gy}\right)\label{eq:Uu equation}\end{equation}
and \begin{equation}
\bm{U}_{X}=\frac{\bm{B}^{\dagger}}{B_{\parallel}^{\dagger}}\bm{U}_{X}\cdot\bm{b}+\frac{c}{eB_{\parallel}^{\dagger}}\bm{b}\times\left(\mu\nabla B+\nabla H_{gy}\right)\label{eq:Ux equation}\end{equation}
when $\bm{B}^{\dagger}\cdot$ and $\bm{b}\times$ are applied respectively.
Similarly, the other $\delta l_{GK}/\delta U^{J}$ equations give
\begin{align}
 & U_{\mu}=0,\label{eq:Umu equation}\\
 & \bm{b}\cdot\bm{U}_{X}=u+\frac{1}{m}\frac{\partial H_{gy}}{\partial u},\label{eq:b.Ux equation}\\
 & U_{\theta}=\frac{eB}{mc}+\frac{e^{2}}{mc^{2}}\bm{U}_{X}\cdot\frac{\partial\bm{A}^{\dagger}}{\partial\mu}+\frac{e}{mc}\frac{\partial H_{gy}}{\partial\mu},\label{eq:U theta equation}\end{align}
and Eq.~\eqref{eq:b.Ux equation} is combined wiht Eq.~\eqref{eq:Ux equation}
to give $\bm{U}_{X}$ in terms of co-ordinates. The form of Eqs.~\eqref{eq:Uu equation}-\eqref{eq:U theta equation}
is identical to the standard Lagrangian equations for $\left(\dot{\bm{X}},\dot{u},\dot{\mu},\dot{\theta}\right)$
because of the linearity of the Lagrangian in $\bm{U}$. Note that
various terms have different origins in the Lagrangian and Eulerian
derivations; for instance, $\partial\bm{X}/\partial t=0$ in the Eulerian
derivation (it is just a co-ordinate), while this is not true in the
Lagrangian case. 

Equipped with the solution for $\bm{U}$ in terms of phase space co-ordinates,
the gyrokinetic Vlasov equation is Eq.~\eqref{eq:GK 6-form advection},
or in co-ordinates, \begin{equation}
\partial_{t}
\hat{F}+\nabla\cdot\left(\bm{U}\hat{F}\right)=0.\label{eq:GK Vlasov equation}\end{equation}
Maxwell's equations follow from the standard Euler-Lagrange equations
for $\bm{A}$ and $\phi$, \begin{equation}
\frac{\delta l_{GK}}{\delta\bm{A}_{1}}=\frac{\delta l_{GK}}{\delta\phi_{1}}=0,\label{eq:Func derivs for Maxwell's eqns}\end{equation}
since $\dot{\phi}_{1}$ and $\dot{\bm{A}}_{1}$ do not appear in $l_{GK}$.
These lead to the gyrokinetic Maxwell's equations for $\phi_{1}$
and $\bm{A}_{1}$, 
\begin{align}
 & \frac{1}{4\pi}\nabla^{2}\phi_{1}\left(\bm{x}\right)=-\frac{\delta\mathcal{H}}{\delta\phi_{1}\left(\bm{x}\right)}\label{eq:phi Maxwells eqn}\\
 & \frac{1}{4\pi}\nabla\times\nabla\times\bm{A}_{1}\left(\bm{x}\right)=-\frac{\delta\mathcal{H}}{\delta\bm{A}_{1}\left(\bm{x}\right)}-\frac{1}{4\pi} \nabla \times \bm{B},\label{eq:A Maxwells eqn}\end{align}
where $\mathcal{H}\equiv\sum_{s}\int d\bm{X}dud\mu d\theta\hat{F}H$.

\section{The Hamiltonian formulation and gyrokinetic Poisson brackets\label{sec:The-Hamiltonian-formulation}}

We now perform the generalized Legendre transform of $l_{GK}$ and
use the corresponding Hamiltonian formulation to construct the Poisson
brackets for the gyrokinetic system. One complication is the degeneracy
in the Lagrangian that arises from the lack of quadratic dependence
on $\bm{U}$, $\dot{\bm{A}}$ and $\dot{\phi}$. This issue is discussed
in detail Refs.~\onlinecite{Cendra98themaxwell-vlasov,Gumral:2010p5807} and those same arguments
apply to the gyrokinetic case. 

For the moment, we formulate a bracket on the space of plasma densities (see Section~\ref{sub:Inclusion-of-electromagnetic})
and carry out a Legendre transform in $\bm{U}$ by defining \begin{equation}
\bm{M}=\frac{\delta l_{GK}}{\delta\bm{U}}.\label{eq:legendre variable definition}\end{equation} 
This type of formulation treats the gyrokinetic Poisson-Amp\`{e}re equations [Eqs.~\eqref{eq:phi Maxwells eqn} and \eqref{eq:A Maxwells eqn}] as \emph{constraints} on the 
motion of $\hat{F}$, rather than dynamical equations in their own right. 
The gyrokinetic Hamiltonian is defined, as for a standard Legendre transform, as\begin{align}
h_{GK} & =\left\langle \bm{M},\bm{U}\right\rangle_\mathfrak{g} -l_{GK}\left(\bm{U},\hat{F}\right)\nonumber \\
= & \sum_{s}\int d\bm{X}dud\mu d\theta\left[\bm{M}\cdot\bm{U}-\hat{F}\left(\frac{e}{c}\bm{A}^{\dagger}\cdot\bm{U}_{X}\right.\right.\nonumber \\
 +&\frac{mc}{e}\left.\left.\mu U_{\theta}-H\right)\right]-\frac{1}{8\pi}\int d\bm{x}\left(\left|\nabla\phi_{1}\right|^{2}-\left|\nabla\times\left(\bm{A}+\bm{A}_{1}\right)\right|^{2}\right).\label{eq:Full GK Hamiltonian}\end{align}
It is easy to show that with this Hamiltonian\begin{align}
\left\{ \Gamma,\Theta\right\} _{LP} & =-\left\langle \bm{M},\left[\frac{\delta\Gamma}{\delta\bm{M}},\frac{\delta\Theta}{\delta\bm{M}}\right]\right\rangle _{\mathfrak{g}}\nonumber \\
 & +\left\langle \hat{F},\frac{\delta\Theta}{\delta\bm{M}}\cdot\nabla\frac{\delta\Gamma}{\delta\hat{F}}-\frac{\delta\Gamma}{\delta\bm{M}}\cdot\nabla\frac{\delta\Theta}{\delta\hat{F}}\right\rangle _{V}\label{eq:Full GK Poisson Bracket}\end{align}
is a valid Poisson bracket (see Sec.~\ref{EP reduction explanation} and Ref.~\onlinecite{Cendra98themaxwell-vlasov}). 
To evaluate functional derivatives of $h_{GK}$ [Eq.~\eqref{eq:Full GK Hamiltonian}], one should 
obtain the Green's function solutions for $\phi_1$ and $\bm{A}_1$, for instance
\begin{equation}
\phi_1\left(\bm{x}\right)=\sum_s\int d\bm{X}'du' d\mu' d\theta' K\left(\bm{x}|\bm{z}'\right) \hat{F}\left(\bm{z}'\right),
\label{eq:phi Greens function}
\end{equation}
from the gyrokinetic Poisson-Amp\`{e}re equations, and insert these into $h_{GK}$, see 
Refs.~\onlinecite{Gumral:2010p5807,1982AIPC...88...13M}. 
For practical calculation, this is the same as neglecting the electromagnetic part of $h_{GK}$ in 
the functional derivative. 
In the same way as the Maxwell-Vlasov system (Sec.~\ref{EP reduction explanation}), the manifestation of the bracket in Eq.~\eqref{eq:Full GK Poisson Bracket} is not well defined due to the constraints on $\bm{M}$. In the next section, the Dirac theory of constraints (see Appxs.~\ref{Appendix DC} and \ref{Appendix ModDC}) 
is used to reduce Eq.~\eqref{eq:Full GK Poisson Bracket} to a bracket of the space of densities $\hat{F}$.

A complete treatment of the geometry of the Poisson-Vlasov
system, with the electric field as a constraint, is given
in Ref.~\onlinecite{Gumral:2010p5807}. Many similar ideas 
will apply to the gyrokinetic system, with complications arising from the 
nonlocal nature of the theory\cite{Qin:2007p5801} and larger constraint space 
($\phi_1$ and $\bm{A}_1$ rather than just $\phi$). We reiterate that there are two 
sets of constraints we consider here; the constraints on $\bm{M}$ variables, similar to the 
Maxwell-Vlasov system, and the constraints due to $\phi_1$ and $\bm{A}_1$, which are the gyrokinetic 
Poisson-Amp\`{e}re equations. We first deal with the $\bm{M}$ constraints, eliminating these variables 
entirely, then explain how to include $\phi_1$ and $\bm{A}_1$ in Section~\ref{sub:Inclusion-of-electromagnetic}.

\subsection{Gyrokinetic Poisson bracket\label{Gyrokinetic Poisson bracket}}

There are six constraints given by \begin{equation}
\Phi_{I}\left(\bm{z}\right)=M_{I}\left(\bm{z}\right)-\frac{\delta l_{GK}}{\delta U^{I}\left(\bm{z}\right)}=0,\label{General constraints}\end{equation}
with the functional derivatives as listed in Eq.~\eqref{eq:lGK functional derivatives}.
One then forms the constraint matrix $\mathcal{C}_{IJ}\left(\bm{z},\bm{z}'\right)=\left\{ \Phi_{I}\left(\bm{z}\right),\Phi_{J}\left(\bm{z}'\right)\right\} $
with the Poisson bracket of Eq.~\eqref{eq:Full GK Poisson Bracket}
using\begin{align}
\frac{\delta\Phi_{I}\left(\bm{z}\right)}{\delta M_{J}\left(\bar{\bm{z}}\right)} & =\delta_{I}^{J}\delta\left(\bm{z}-\bm{\bar{z}}\right)\delta_{ss'},\:\frac{\delta\Phi_{u}\left(\bm{z}\right)}{\delta\hat{F}\left(\bar{\bm{z}}\right)}=\frac{\delta\Phi_{\mu}}{\delta\hat{F}}=0,\nonumber\\
\frac{\delta\Phi_{i}\left(\bm{z}\right)}{\delta\hat{F}\left(\bar{\bm{z}}\right)} & =-\frac{e}{c}A_{i}^{\dagger}\left(\bar{\bm{z}}\right)\delta\left(\bm{z}-\bar{\bm{z}}\right)\delta_{ss'},\nonumber\\
\frac{\delta\Phi_{\theta}\left(\bm{z}\right)}{\delta\hat{F}\left(\bar{\bm{z}}\right)} & =-\bar{\mu}\,\delta\left(\bm{z}-\bar{\bm{z}}\right)\delta_{ss'}.\label{VarDerivs of constraints}\end{align}
Dropping boundary terms in integrations and inserting the constraint
equations (after calculation of the brackets, see Appx.~\ref{Appendix ModDC}) leads to the very simple
form,\begin{align}
 & \mathcal{C}_{IJ}\left(\bm{z},\bm{z}'\right)=\hat{F}\delta\left(\bm{z}-\bm{z}'\right)\delta_{ss'}\nonumber \\
 &\times \left(\begin{array}{cccccc}
0 & -\frac{e}{c}B_{z}^{\dagger} & \frac{e}{c}B_{y}^{\dagger} & -mb_{x} & \frac{mc}{e}W_{x} & 0\\
\frac{e}{c}B_{z}^{\dagger} & 0 & -\frac{e}{c}B_{x}^{\dagger} & -mb_{y} & \frac{mc}{e}W_{y} & 0\\
-\frac{e}{c}B_{y}^{\dagger} & \frac{e}{c}B_{x}^{\dagger} & 0 & -mb_{z} & \frac{mc}{e}W_{z} & 0\\
mb_{x} & mb_{y} & mb_{z} & 0 & 0 & 0\\
-\frac{mc}{e}W_{x} & -\frac{mc}{e}W_{y} & -\frac{mc}{e}W_{z} & 0 & 0 & -\frac{mc}{e}\\
0 & 0 & 0 & 0 & \frac{mc}{e} & 0\end{array}\right),\label{eq:CIJ matrix}\end{align}
where all functions are of the $\bm{z}$ variable and $ $$\bm{W}=\bm{R}+\frac{1}{2}\bm{b}\,\bm{b}\cdot\nabla\times\bm{b}$.
Because of the simple form in $\bm{z}'$, this matrix is 
easy to invert according to Eq.~\eqref{eq:Dirac Cinv defn} giving,
\begin{align}
 & \mathcal{C}_{IJ}^{-1}\left(\bm{z},\bm{z}'\right)=\frac{1}{\hat{F}B_{\parallel}^{\dagger}}\delta\left(\bm{z}-\bm{z}'\right)\delta_{ss'}\nonumber \\
 & \times\left(\begin{array}{cccccc}
0 & \frac{c}{e}b_{z} & -\frac{c}{e}b_{y} & \frac{1}{m}B_{x}^{\dagger} & 0 & \frac{c}{e}\hat{W}_{x}\\
-\frac{c}{e}b_{z} & 0 & \frac{c}{e}b_{x} & \frac{1}{m}B_{y}^{\dagger} & 0 & \frac{c}{e}\hat{W}_{y}\\
\frac{c}{e}b_{y} & -\frac{c}{e}b_{x} & 0 & \frac{1}{m}B_{z}^{\dagger} & 0 & \frac{c}{e}\hat{W}_{z}\\
-\frac{1}{m}B_{x}^{\dagger} & -\frac{1}{m}B_{y}^{\dagger} & -\frac{1}{m}B_{z}^{\dagger} & 0 & 0 & \frac{1}{m}W^{\dagger}\\
0 & 0 & 0 & 0 & 0 & \frac{e}{mc}B_{\parallel}^{\dagger}\\
-\frac{c}{e}\hat{W}_{x} & -\frac{c}{e}\hat{W}_{y} & -\frac{c}{e}\hat{W}_{z} & -\frac{1}{m}W^{\dagger} & -\frac{e}{mc}B_{\parallel}^{\dagger} & 0\end{array}\right),\label{eq:invCIJ matrix}\end{align}
where again functions are of the $\bm{z}$ variable, $\hat{\bm{W}}\equiv\bm{b}\times\bm{W}$
and $W^{\dagger}\equiv\bm{B}^{\dagger}\cdot\bm{W}$. Of course, this
matrix is nothing but the single particle gyrokinetic Poisson matrix\cite{Brizard:2007p5792} as was
the case for the Maxwell-Vlasov system.
Restricting the functionals $\Gamma$ and $\Theta$ to not depend on $\bm{M}$ 
(see Appx.~\ref{Appendix ModDC}) and using \begin{equation}
\left\{ \Gamma[\hat{F}],\Phi_{J}\left(\bm{z}\right)\right\} =\hat{F}\left(\bm{z}\right)\frac{\partial}{\partial z^{J}}\frac{\delta\Gamma}{\delta\hat{F}},\label{eq:PB with Gamma and Phi}\end{equation}
the field theory gyrokinetic Poisson bracket is simply,\begin{equation}
\left\{ \Gamma,\Theta\right\} _{DB}=\left\langle \hat{F},\left\{ \frac{\delta\Gamma}{\delta\hat{F}},\frac{\delta\Theta}{\delta\hat{F}}\right\} _{sp}\right\rangle _{V}.\label{eq:GK PB, no potentials}\end{equation}
Here $\left\{ \,,\right\} _{sp}$ is the single particle Poisson
bracket structure\cite{CorreaRestrepo:2005p10307,Brizard:2007p5792}
\begin{align}
\left\{ f,g\right\} _{sp} & =-\frac{c\bm{b}}{eB_{\parallel}^{\dagger}}\cdot\nabla f\times\nabla g+\frac{\bm{B}^{\dagger}}{mB_{\parallel}^{\dagger}}\cdot\left(\nabla f\frac{\partial g}{\partial u}-\nabla g\frac{\partial f}{\partial u}\right)\nonumber \\
+\frac{c\hat{\bm{W}}}{e} & \cdot\left(\nabla f\frac{\partial g}{\partial\theta}-\nabla g\frac{\partial f}{\partial\theta}\right)+\frac{W^{\dagger}}{m}\left(\frac{\partial f}{\partial u}\frac{\partial g}{\partial\theta}-\frac{\partial g}{\partial u}\frac{\partial f}{\partial\theta}\right)\nonumber \\
+\frac{e}{mc} & \left(\frac{\partial f}{\partial\mu}\frac{\partial g}{\partial\theta}-\frac{\partial f}{\partial\theta}\frac{\partial g}{\partial\mu}\right).\label{eq:Single part GK PB}\end{align}
We note that, although all $\delta/\delta \bm{M}$ terms are left out above for clarity, the $\delta/\delta M_u$ and $\delta/\delta M_{\mu}$ terms cancel in a full calculation as expected, so there is no issue with these being undefined (see Appx.~\ref{Appendix ModDC}). 
The field theory bracket, Eq.~\eqref{eq:GK PB, no potentials}, is
of exactly the form one would expect based on the Poisson-Vlasov bracket\cite{1982AIPC...88...13M} and Maxwell-Vlasov bracket\cite{Marsden:1982p9058,Morrison:1980:VMbracket} (Eq.~\eqref{eq:Full VM bracket} without $\delta/\delta\bm{E}$
and $\delta/\delta\bm{B}$ terms). It is aesthetically pleasing to see this type of structure emerge from the
entirely systematic procedure applied above. Since we have not given a
proof of the modified Dirac procedure used in the calculation (see Appx.~\ref{Appendix ModDC}), we directly prove 
the Jacobi identity for the gyrokinetic bracket [Eq.~\eqref{eq:GK PB, no potentials}] in Appx.~\ref{Appendix JI Proof}.
This shows it is indeed a valid Poisson bracket for this gyrokinetic system. 
The reduced Hamiltonian
to be used with Eq.~\eqref{eq:GK PB, no potentials} is simply Eq.~\eqref{eq:Full GK Hamiltonian}
with constraints on $\bm{M}$ inserted explicitly,\begin{equation}
h=\sum_{s}\int d\bm{X}dud\mu d\theta\hat{F}H-\frac{1}{8\pi}\int d\bm{x}\left(\left|\nabla\phi_{1}\right|^{2}-\left|\nabla\times\left(\bm{A}+\bm{A}_{1}\right)\right|^{2}\right).\label{eq:Reduced Hamiltonian, no potentials}\end{equation}
It is easy to show that $\partial_{t}\hat{F}=\left\{ \hat{F},h\right\} $
is just the conservative form of the gyrokinetic Vlasov equation,
Eq.~\eqref{eq:GK Vlasov equation}.

\subsection{Inclusion of electromagnetic fields\label{sub:Inclusion-of-electromagnetic}}

The bracket, Eq.~\eqref{eq:Single part GK PB}, does not include
electromagnetic field equations, meaning the gyrokinetic Maxwell's equations, Eqs.~\eqref{eq:phi Maxwells eqn}
and \eqref{eq:A Maxwells eqn}, must be specified as separate constraints on the motion to obtain
a closed system. Here, we illustrate how to explicitly include the electromagnetic potentials in 
the bracket for a simplified gyrokinetic
system. This procedure also works to extend the simple Poisson-Vlasov
bracket\cite{1982AIPC...88...13M,Gumral:2010p5807} to include the
motion of $\phi$.  The general technique is to add a Poisson-Amp\`{e}re canonical bracket
to the gyrokinetic bracket [Eq.~\eqref{eq:Single part GK PB}] and apply standard Dirac theory 
to this extended bracket. It is important to 
recognize that this is only valid because the full constraint matrix would be block 
diagonal if the reduction were performed in one-step from an original bracket that included electromagnetic and 
plasma components (i.e., Eq.~\eqref{eq:Full GK Poisson Bracket} with the addition of canonical brackets in 
$\bm{A}_1$ and $\phi_1$). This condition is satisfied because $\bm{A}_1$ and $\phi_1$ do not appear in the symplectic 
structure of the original Lagrangian. 

For clarity, we use with a simplified electrostatic system in the
drift kinetic limit, with $H=e\phi_{1}+m\left|\delta\bm{u}_{E}\right|^{2}/2$
where $\delta\bm{u}_{E}=c\left(\bm{b}\times\nabla\phi_{1}\right)/B$.
We also assume quasineutrality, which amounts to neglecting the $\int d\bm{x}\left|\nabla\phi\right|^{2}/8\pi$
term in the Lagrangian, and set $\bm{W}$ to zero%
\footnote{With $\bm{W}=0$ the gyrokinetic Lagrangian is no longer gyrogauge
invariant, an issue of little practical importance for formulating
such a model.%
}. The Hamiltonian for the system is \begin{align}
h= & \int d\bm{X}\phi\left(\bm{X}\right)\Pi\left(\bm{X}\right)\nonumber \\
+ & \sum_{s}\int d\bm{X}dud\mu d\theta\hat{F}\left(\frac{m}{2}u^{2}+\mu B+e\phi+\frac{mc^{2}}{2B^{2}}\left|\nabla_{\perp}\phi\right|^{2}\right),\label{eq:Simplified unreduced GK h}\end{align}
and the unreduced Poisson bracket\begin{equation}
\left\{ \Gamma,\Theta\right\} =\left\langle \hat{F},\left\{ \frac{\delta\Gamma}{\delta\hat{F}},\frac{\delta\Theta}{\delta\hat{F}}\right\} _{sp}\right\rangle _{V}+\int d\bm{X}\left(\frac{\delta\Gamma}{\delta\phi}\frac{\delta\Theta}{\delta\Pi}-\frac{\delta\Theta}{\delta\phi}\frac{\delta\Gamma}{\delta\Pi}\right),\label{eq:Simplified unreduced GK PB}\end{equation}
where $\Pi=\delta l/\delta\dot{\phi}$ is the variable canonically
conjugate to $\phi$. This model is the electrostatic version of the
simplified gyrokinetic system in Refs.~\onlinecite{Scott:2010p9381,Scott:2010p9978}. Physically, the
$m\left|\delta\bm{u}_{E}\right|^{2}/2=mc^2/2B^2 \left|\nabla_{\perp}\phi\right|^{2}$ term in the Hamiltonian is
the polarization drift in the drift kinetic limit\cite{Scott:2010p9978,Dubin:1983p9400}. 
$\nabla_{\perp}$ indicates a gradient with respect to a co-ordinate system locally perpendicular 
to the background magnetic field (we are neglecting derivatives of $\bm{b}$). Unlike in the previous 
section, $\phi$ is now considered a separate field in the Hamiltonian, and Poisson's 
equation should not be used to evaluate functional derivatives.

The constraints are \begin{align}
\Phi_{1} & =\frac{\delta h}{\delta\phi}=\sum_{s}\int dud\mu d\theta\left[e\hat{F}-mc^{2}\nabla_{\perp}\cdot\left(\frac{\hat{F}}{B^{2}}\nabla_{\perp}\phi\right)\right],\nonumber \\
\Phi_{2} & =\Pi,\label{eq:PB with phi constraints}\end{align}
where $\Pi=0$ since $\delta l/\delta\dot{\phi}=0$. $\Phi_{1}=0$
is the gyrokinetic Poisson equation; this constraint arises as a secondary Dirac constraint that is necessary to satisfy $\dot{\Phi}_{2}=0$,
see Ref.~\onlinecite{Chandre:2011p9324} for more information. Using Eq.~\eqref{eq:Simplified unreduced GK PB}
the constraint matrix is, \begin{equation}
\mathcal{C}_{IJ}\left(\bm{X},\bm{X}'\right)=\left(\begin{array}{cc}
0 & C\\
-C & 0\end{array}\right),\label{eq:Constraint matrix potentials}\end{equation}
where \begin{equation}
C=\frac{\delta\Phi_{1}\left(\bm{X}\right)}{\delta\phi\left(\bm{X}'\right)}=-c^{2}\nabla'_{\perp}\cdot\left[\frac{\hat{n}\left(\bm{X}'\right)}{B^{2}\left(\bm{X}'\right)}\nabla'_{\perp}\delta\left(\bm{X}-\bm{X}'\right)\right]\label{eq:Func deriv for constraint mat}\end{equation}
 with $\hat{n}=\sum_{s}\int dud\mu d\theta m\hat{F}$ and $\nabla'_{\perp}$
indicating the derivative is with respect to $\bm{X}'_{\perp}$. The
inverse matrix, chosen to satisfy Eq.~\eqref{eq:Dirac Cinv defn},
is \begin{equation}
\mathcal{C}_{IJ}^{-1}\left(\bm{X},\bm{X}'\right)=\left(\begin{array}{cc}
0 & -C^{-1}\\
C^{-1} & 0\end{array}\right),\label{eq:InvC with potentials}\end{equation}
where \begin{equation}
C^{-1}\left(\bm{X},\bm{X}'\right)=-\frac{1}{c^{2}}\nabla_{\perp}^{-1}\cdot\left[\frac{B^{2}\left(\bm{X}\right)}{\hat{n}\left(\bm{X}\right)}\nabla_{\perp}^{-1}\delta\left(\bm{X}-\bm{X}'\right)\right].\label{eq:func deriv for Cinv mat}\end{equation}
The notation ${\nabla_{\perp}^{-1}\cdot\left[B^{2}\left(\bm{X}\right)/\hat{n}\left(\bm{X}\right)\nabla_{\perp}^{-1}\:\cdot\:\right]}$ is used for clarity to denote the inverse of the operator ${\nabla_{\perp}\cdot\left[\hat{n}\left(\bm{X}\right)/B^{2}\left(\bm{X}\right)\nabla_{\perp}\:\cdot\:\right]}$, which is similar to the perpendicular Laplacian. While 
we will not consider this here, existence of the inverse of this type of operator (with smooth $\hat{n}/B^2$)
can very likely be proven 
using standard partial differential equation analysis techniques\cite{BersPDEs1957}.
The Dirac bracket is constructed using\begin{align}
\left\{ \Gamma,\Phi_{1}\left(\bm{X}\right)\right\}  & =c^{2}\nabla_{\perp}\cdot\left(\frac{\hat{n}}{B^{2}}\nabla_{\perp}\frac{\delta\Gamma}{\delta\Pi}\right)\nonumber \\
+ & \tilde{N}_{\Gamma}+\frac{mc^{2}}{e}\nabla_{\perp}\cdot\left(\frac{\nabla_{\perp}\phi}{B^{2}}\tilde{N}_{\Gamma}\right),\nonumber \\
\left\{ \Gamma,\Phi_{2}\left(\bm{X}\right)\right\}  & =\frac{\delta\Gamma}{\delta\phi\left(\bm{X}\right)},\label{eq:Brackets with Gamma for DB}\end{align}
where \begin{equation}
\tilde{N}_{\Gamma}=\sum_{s}\int dud\mu d\theta\,\frac{1}{m}\nabla\cdot\left(cf\bm{b}\times\nabla\frac{\delta\Gamma}{\delta\hat{F}}+\frac{e}{m}f\bm{B}^{\dagger}\frac{\partial}{\partial u}\frac{\delta\Gamma}{\delta\hat{F}}\right),\label{eq:Ntilde definition for PB}\end{equation}
with the corresponding definition for $\tilde{N}_{\Theta}$. With
Eq.~\eqref{eq:Dirac Brak Defn}, this leads to
 \begin{align}
\left\{ \Gamma,\Theta\right\} _{DB}= \sum_{s}\int & d\bm{X}dud\mu d\theta\hat{F}\left\{ \frac{\delta\Gamma}{\delta\hat{F}},\frac{\delta\Theta}{\delta\hat{F}}\right\} _{sp}\nonumber \\
+\int d\bm{X}\nabla_{\perp}^{-1} & \cdot\left(\frac{B^{2}}{c^{2}\hat{n}}\nabla_{\perp}^{-1}\frac{\delta\Theta}{\delta\phi}\right)\left[\tilde{N}_{\Gamma}+\frac{mc^{2}}{e}\nabla_{\perp}\cdot\left(\frac{\nabla_{\perp}\phi}{B^{2}}\tilde{N}_{\Gamma}\right)\right]\nonumber \\
-\int d\bm{X}\nabla_{\perp}^{-1} & \cdot\left(\frac{B^{2}}{c^{2}\hat{n}}\nabla_{\perp}^{-1}\frac{\delta\Gamma}{\delta\phi}\right)\left[\tilde{N}_{\Theta}+\frac{mc^{2}}{e}\nabla_{\perp}\cdot\left(\frac{\nabla_{\perp}\phi}{B^{2}}\tilde{N}_{\Theta}\right)\right].\label{eq:Dirac bracket with potentials}
\end{align}
Here, $\left\{ \,,\right\} _{sp}$ is the single particle bracket
as in Eq.~\eqref{eq:Single part GK PB} (with $\bm{W}=0$). Note
that in forming Eq.~\eqref{eq:Dirac bracket with potentials}, terms
involving $\delta/\delta\Pi$ in the Dirac part of Eq.~\eqref{eq:Dirac Brak Defn},
canceled with the canonical part of the original bracket, as would
be expected. 

With the reduced Hamiltonian (Eq.~\eqref{eq:Simplified unreduced GK h}
without the first term), the bracket can easily be checked to give
the Vlasov equation as $\partial_{t}\hat{F}\left(\bm{X}\right)=\left\{ \hat{F}\left(\bm{X}\right),h\right\} _{DB}$.
Noticing that the $\partial/\partial u$ term in $\partial_{t}\hat{F}$
{[}see Eq.~\eqref{eq:GK Vlasov equation}{]} integrates to zero,
we see that $\sum_{s}\int dud\mu d\theta\, e\partial_{t}\hat{F}=-\tilde{N}_{h}$.
This is used in $\partial_{t}\phi\left(\bm{X}\right)=\left\{ \phi\left(\bm{X}\right),h\right\} _{DB}$
to show\begin{equation}
\sum_{s}\int dud\mu d\theta\left\{ e\partial_{t}\hat{F}-mc^{2}\nabla_{\perp}\cdot\left[\frac{1}{B^{2}}\frac{\partial}{\partial t}\left(\hat{F}\nabla_{\perp}\phi\right)\right]\right\} =0,\label{eq:Time derivative of Poisson's}\end{equation}
which is just the time derivative of Poisson's equation for this gyrokinetic
model. Using the procedure presented above there should be no particular
obstacle to the construction of brackets for more complex gyrokinetic
theories. For instance, one could include finite Larmor radius effects
or magnetic fluctuations%
\footnote{Including full magnetic fluctuations would require careful treatment
of the electromagnetic gauge. This could be handled using first class
constraints for unphysical degrees of freedom\cite{Dirac:1950p9409}
or by fixing the gauge with a Lagrange multiplier, $\lambda$, and
treating $\lambda$ as a separate variable\cite{Sugama:2000p9020}.%
}. However, considering the complexity of the bracket for even a very
simple gyrokinetic model, such brackets are unlikely to be of much
practical use.

\section{Numerical applications\label{sec:Numerical-applications}}

In a general sense, one of the main motivations for this work is to possibly help 
address the increasing need to develop new algorithms with the long-time 
conservation properties necessary to help improve the 
physics fidelity of simulation results as we move towards  
exascale computing and beyond. Of specific interest in this paper is the desire to explore the possibility of 
applying the Eulerian variational methods developed here in a  \emph{discrete} 
context to design continuum geometric
integrators for gyrokinetic systems. To elaborate on this idea, here
we give a simple explanation of some geometric discretization methods
based on recent work in numerical fluid dynamics. The methods described
here are just examples from a large array of literature on the subject.
Some other techniques can be found in, for instance Refs.~\onlinecite{Cotter2007,Elcott:2007:SCS:1189762.1189766,Mullen:2009:EIF:1531326.1531344,Zhang2002764,Bridges:2006p9373}.
In addition we remark on how Euler-Poincar\'{e} models can be used to
formulate sub-grid models for turbulence simulation and some of the
challenges associated with extending these ideas to gyrokinetic turbulence.

\paragraph{Lagrangian side: discrete Euler-Poincar\'{e} equations}

Conceptually, an obvious way to design a geometric integrator for
an Euler-Poincar\'{e} system is to directly discretize the Euler-Poincar\'{e}
variational principle. If one can design discrete variations of the
correct form, the entire integrator can be constructed directly from
the variational principle as for a standard variational integrator.
This approach has recently been successfully applied to develop an
integrator for the Euler fluid equations\cite{Pavlov:2009p5951} and
more complex fluids, including magnetohydrodynamics (MHD)\cite{Gawlik:2011p5814}.
The utility of such an approach is illustrated by the very nice properties
of these schemes. For instance, the MHD scheme\cite{Gawlik:2011p5814}
exactly preserves $\nabla\cdot\bm{B}=0$ and the cross helicity $\int d\bm{x}\,\bm{v\cdot\bm{B}}$.
As one consequence of this, there is almost no artificial magnetic
reconnection. The symplectic nature
of the scheme also leads to other very good long time conservation properties.

The first requirement in constructing an Euler-Poincar\'{e} integrator
is a finite dimensional approximation to the diffeomorphism Lie group.
In the case of fluids or MHD, the group is that of \emph{volume preserving}
diffeomorphisms and a matrix Lie group is constructed to satisfy analogous
properties to the infinite dimensional group. For Vlasov-Poisson,
Vlasov-Maxwell or a gyrokinetic system, the group is that of\emph{
symplectomorphisms}. Thus, for a discretization, a different matrix
Lie group than the fluid case should be used, with properties designed
to mimic those of the infinite dimensional symplectomorphism group.
Using this group one can find the Lie algebra, which will give the
form of the space of discrete vector fields (just the Eulerian phase
space fluid velocities). Group operations can then be constructed
as matrix multiplications as for a standard finite dimensional Lie
group, and advected parameters included through the use of discrete
exterior calculus. One would then use the \emph{discrete Euler-Poincar\'{e}
theorem}\cite{BouRabee:2009p7780}, which gives discrete update equations
{[}in analogy with Eq.~\eqref{eq:Full EP deivation}{]} from a discrete
reduced Lagrangian. An algorithm of this form can be shown to be symplectic
and have similar conservation properties (arising from variants of
Noether's theorem) to the continuous system. The final update equations
obtained from this method are not as complex as one might expect and
would not preclude incorporation into large scale codes. Obviously
there are several unanswered questions regarding the application of
this method to kinetic plasma systems. First,  one must discretize the symplectomorphism group,
which may not be trivial. In two phase space dimensions the group is the same as the group of
volume preserving diffeomorphisms; however, in higher dimensions the symplectomorphisms 
form a more restricted class of transformations. 
The lack of a finite boundary in velocity space  may also present issues relating to the discretization of the symplectomorphisms. The degeneracy of the system is another aspect which
differs from the fluid system, and the consequences of this in the
discrete setting would have to be carefully considered. Finally, for
a gyrokinetic system, it would be necessary to remove the $\theta$
dimension in some way. This could potentially be done either in the
continuous setting or after discrete equations have been obtained.

\paragraph{Hamiltonian side: Poisson bracket discretization}

Another way to form a discrete Hamiltonian system is to directly discretize
the Poisson bracket. The general idea is simple, one finds a discrete
Hamiltonian functional and discrete bracket that are finite dimensional
approximations to the continuous versions. In this way, one discretizes
(in phase space) via the method of lines, and reduces the infinite
dimensional system to an approximate finite dimensional one. Any symplectic
temporal discretization can then be used to ensure the system is discretely
Hamiltonian \cite{Bridges:2006p9373}. The difficultly arises in ensuring
a correct Hamiltonian discretization of the bracket. This requires
antisymmetry and the Jacobi identity to be satisfied, and such a bracket
can be very difficult to find in practice. For instance, for the Euler
fluid equations, the non-canonical structure complicates matters and
a discrete bracket has been found only for simplified cases\cite{McLachlan:1993p9418}.
An obvious place to start in this endeavor would be the Vlasov-Poisson
system, as the structure is much more simple. Generalizations to gyrokinetic
systems could then potentially be achieved through Nambu bracket formulations\cite{Scott:2010p9381}.

\paragraph{Alpha models and large-eddy simulation}

Much work has been done in the last decade in the fluids community
on so-called \emph{alpha models. }The general idea is to regularize
the fluid equations (Navier-Stokes or MHD) at the level of the Euler-Poincar\'{e}
variational principle, by adding terms into the Lagrangian that include
gradients of the fluid velocity. These terms penalize the formation
of small scale structures, and can thus be used as a large eddy simulation
(LES) model, causing turbulence to dissipate at larger scales\cite{Chen:1998p9384}.
These methods have been shown to have some significant advantages over
more traditional LES methods (for instance those based on hyperdiffusion)
especially for simulation of MHD turbulence\cite{Chen:1999p9395,PietarilaGraham:2011p9392}.
As gyrokinetic turbulence simulation becomes a more mature subject,
it is interesting to enquire whether similar alpha models could be
formulated for gyrokinetic large eddy simulation. 

In fact, alpha models can be \emph{derived }from a standard fluid
variational principle, by averaging over small scale fluctuations
that are assumed to be advected by the larger scale flow\cite{Holm:2002p8805,Bhat:2005p9397}.
Approaching the gyrokinetic variational principle in a similar way
leads to the addition of extra, regularized terms into the gyrokinetic
Lagrangian. For instance, following the general ideas in Ref.~\onlinecite{Bhat:2005p9397},
averaging over perpendicular $\bm{X}$-space fluctuations of scale
length $\alpha$, and ensuring electromagnetic gauge invariance, we were led to the
regularized Lagrangian $l=l_{GK}+l_{\alpha}$, where \begin{equation}
l_{\alpha}=\alpha^{2}\int d\bm{X}dud\mu d\theta\hat{F}\left(\bm{B}^\dagger \cdot\nabla\times\bm{U}_{X}-\nabla_{\perp}^{2}H\right).\label{eq:Regularized GK lagrangian}\end{equation}
While this Lagrangian gives well-defined equations of motion, there
is a fundamental problem in that it destroys some of the degeneracy
in the original system. As a consequence of this, the equations of
motion involve solving spatial PDEs for $\bm{U}$, which would significantly
increase computation times, defeating the purpose of an LES. It is not 
yet clear if it is possible to design a regularization of this type for the gyrokinetic system 
that retains the redundancy of the $\bm{U}$ fields and allows one to write down 
a standard Vlasov equation. 
We note that gyrokinetic LES has been explored and implemented recently
on the GENE code, by adding hyperdiffusive terms in the perpendicular
co-ordinates\cite{Morel:2011p9388,Morel2012PhPl...19a2311M}.

\section{Concluding Remarks\label{sec:Concluding-Remarks}}

In this article we have applied the Euler-Poincar\'{e} formalism to derive
a new gyrokinetic action principle in Eulerian co-ordinates. We start
with a single-particle Poincar\'{e}-Cartan 1-form, using the theory of
Ref.~\onlinecite{Qin:2007p5801} to systematically construct a gyrokinetic
field theory action in Lagrangian co-ordinates. The fundamental idea
is then to reduce this action using symmetry under the the particle-relabeling map, $\left(\bm{x},\bm{v}\right)=\psi\left(\bm{x}_{0},\bm{v}_{0}\right)$,
which takes particles with initial position $\left(\bm{x}_{0},\bm{v}_{0}\right)$
to their current location $\left(\bm{x},\bm{v}\right)$\cite{holm2009geometric,Cendra98themaxwell-vlasov,Holm:1998p9060}.
This process leads to an action functional formulated in terms of
the Eulerian phase space fluid velocity, $\bm{U}$, and the advected
plasma density, $\hat{F}$, as well as the standard electromagnetic
potentials. In the course of reduction, the arbitrary variations of
the Lagrangian fields (used to derive equations of motion) lead to
\emph{constrained} variations of the Eulerian fields, $\bm{U}$ and
$\hat{F}$. Because of this, field motion is governed by the \emph{Euler-Poincar\'{e}
}equations rather than the standard Euler-Lagrange equations. Explicit
calculation of the Euler-Poincar\'{e} equations for a standard gyrokinetic
single particle Lagrangian is shown to give the gyrokinetic Vlasov
equation. Since the space of electromagnetic potentials is not altered by $\psi\left(\bm{x}_{0},\bm{v}_{0}\right)$,
the gyrokinetic Poisson-Amp\`{e}re equations arise from the standard Euler-Lagrange equations for the perturbed potentials. 

Using the methodology set out in Ref.~\onlinecite{Cendra98themaxwell-vlasov}
we then perform a Legendre transform to derive the Hamiltonian form
of the gyrokinetic system. The principal difficulty is the strong
degeneracy, which is related to the linearity and lack of time derivatives
for certain function variables in the action principle. Physically,
this arises from the fact that the plasma distribution function encodes
the information about particle phase space trajectories. The degeneracy
leads to a Poisson bracket in terms of too many variables; namely,
a series of constrained momentum variables canonically conjugate to
$\bm{U}$ as well as the distribution function $\hat{F}$. To reduce
the bracket into a well defined form we use a modified version of the Dirac theory of constraints,
which is a systematic way to project a Poisson bracket onto a constraint
submanifold when momentum variables are constrained. The modified Dirac 
procedure (see Appx.~\ref{Appendix ModDC}) can be a significant simplification
over standard Dirac theory for certain types of systems. This process leads to
an infinite dimensional gyrokinetic Poisson bracket, which takes a 
natural form based on the single particle bracket. We also demonstrate 
how this procedure leads to the full, electromagnetic 
Vlasov-Maxwell bracket\cite{Marsden:1982p9058,Morrison:1980:VMbracket,Morrison:1980p10549}.
Since the electromagnetic equations in the gyrokinetic system
are really constraints on the motion, we chose to include these in
the bracket via a second application of the Dirac theory of constraints.
The general method is expounded through construction of the bracket
for a simplified electrostatic model with no finite Larmor radius
effects. Although the brackets obtained by such an approach are probably
to complicated to be of much practical use, it makes for an interesting
application of Dirac theory.

As a final comment, we note that there is an emerging (and very likely increasing) demand for a new class of algorithms capable of dealing with the demands of powerful modern supercomputers 
 -- moving on the path to the exascale and beyond\cite{RRosner_Computing2010}.  
 Approaches based on geometric integration may prove highly relevant 
as increasingly ambitious simulations of larger and more complex systems are undertaken.

\section*{Acknowledgements}

We extend our thanks to Dr.~John Krommes and Joshua Burby for 
enlightening discussion. This research is supported by U.S.~DOE (DE-AC02-09CH11466). CC acknowledges financial support from the Agence Nationale de la Recherche and from the European Community under the contract of Association between EURATOM, CEA, and the French Research Federation for fusion studies.


\appendix
\section{Dirac Constraints\label{Appendix DC}}

The Dirac theory of constraints or Dirac bracket is used to build
Poisson brackets for Hamiltonian systems with constraints. The original
purpose of the theory was to construct quantizable Poisson brackets
starting with a degenerate Lagrangian, i.e., a Lagrangian where the
momenta are not independent functions of velocities. The theory applies
equally well to a bracket that is already in non-canonical form, a
realization that can be very useful in the construction of field theoretic
brackets\cite{Morrison20091747,Chandre:2010p9238}. For example, in
Ref.~\onlinecite{Chandre:2011p9324}, the non-canonical magnetohydrodynamic
bracket is reduced to incorporate the incompressibility constraint.
We give a very brief overview of the theory here for the convenience
of the reader. More complete treatments can be found in Refs.~\onlinecite{Chandre:2010p9238,Chandre:2011p9324,marsden1999introduction,Dirac:1950p9409,Morrison20091747}. 

We consider an infinite dimensional Hamiltonian system, described by field variables $\chi_i({\bm z})$ (for $i=1,\ldots,n$), with Poisson
bracket $\left\{ \,,\right\} $, Hamiltonian $H$, and $m$ local constraint
functions $\Phi_{1}({\bm z}),\,\Phi_{2}({\bm z}),\ldots,\Phi_{N}({\bm z})=0$. The constraint
matrix,
\begin{equation}
\mathcal{C}_{ij}\left(\bm{z},\bm{z}'\right)=\left\{ \Phi_{i}({\bm z}),\Phi_{j}({\bm z}')\right\} ,\label{eq:Dirac constraint C}
\end{equation}
and its inverse, defined using \begin{equation}
\int d\bm{z}'\mathcal{C}_{ij}\left(\bm{z},\bm{z}'\right)\mathcal{C}_{jk}^{-1}\left(\bm{z}',\bm{z}''\right)=\delta_{ik}\delta\left(\bm{z}-\bm{z}''\right),\label{eq:Dirac Cinv defn}\end{equation}
are used to form the Dirac bracket,\begin{align}
\left\{ \Gamma,\Theta\right\} _{DB} & =\left\{ \Gamma,\Theta\right\} \nonumber \\
- & \int d\bm{z}d\bm{z}'\left\{ \Gamma,\Phi_{i}\left(\bm{z}\right)\right\} \mathcal{C}_{ij}^{-1}\left(\bm{z},\bm{z}'\right)\left\{ \Phi_{j}\left(\bm{z}'\right),\Theta\right\} .\label{eq:Dirac Brak Defn}\end{align}
By construction this bracket satisfies
the Jacobi identity\cite{Chandre:2010p9238,Chandre:2011p9324,marsden1999introduction,Dirac:1950p9409}. 
Geometrically, the constraints force motion to lie on a \emph{constraint
submanifold, }which inherits the Dirac bracket from the Poisson bracket
on the original manifold\cite{marsden1999introduction}. More precisely, the constraints $\Phi$ are Casimir invariants of the Dirac bracket~(\ref{eq:Dirac Brak Defn}), i.e., $\{\Gamma, \Phi_i\}_{DB}=0$ for any functional $\Gamma$ and for all $j=1,\ldots,m$. 

In the case where the matrix $\mathcal{C}$ is not invertible, Dirac
theory suggests the use of one or more secondary constraints, which must be included
and the constraint matrix $C$ recalculated. See Refs.~\onlinecite{Chandre:2011p9324,marsden1999introduction}
for more information.


\section{Modified Dirac procedure\label{Appendix ModDC}}

In this Appendix we provide some justification of the
modified Dirac procedure used in the calculation of the Vlasov-Maxwell and gyrokinetic
Poisson brackets {[}Eqs.~\eqref{eq:Full VM bracket} and \eqref{eq:GK PB, no potentials}{]}. For clarity we restrict ourselves to the finite dimensional case, even though the technique is applied to infinite dimensional systems in the present manuscript. The purpose of this modified procedure is to simplify the computation of the Dirac bracket and reduce the dimensionality of the constrained system.

The modified Dirac procedure involves the following steps:
\begin{enumerate}
\item Compute the constraint matrix $C$ and simplify its expression by applying the constraints. The resulting matrix $\bar{C}$ is weakly equal to $C$. 
\item Construct the Dirac bracket using Eq.~(\ref{eq:Dirac Brak Defn}) with $\bar{C}$ instead of $C$. 
\item Choose a set of $m$ variables, denoted $\bar{\bm{z}}$, to be eliminated from the bracket~(\ref{eq:Dirac Brak Defn}) using the constraint equations, so 
that it can be completely rewritten in terms of the remaining $n-m$ variables, denoted $\tilde{\bm z}$. Substitute these into the bracket and drop all partial derivatives with respect to the $\bar{\bm{z}}$. This new bracket (of reduced dimensionality) is the \emph{reduced Dirac bracket}. 
\end{enumerate}
Since the above procedure departs from the standard one for the computation of the Dirac bracket, the reduced bracket is not \emph{a priori} a Poisson bracket. Below we justify this procedure by providing the condition under which the reduced Dirac bracket is a Poisson bracket.

Regarding Step 1, assuming $C$ is analytic in the constraint variables, the invertibility of 
the matrix $\bar{C}$ is sufficient to ensure that the reduced Dirac bracket exists, 
in particular, that it satisfies the Jacobi identity.
This follows from the fact that if $\det \bar{C} = \det\left[C\left(\bm{\Phi}=0\right)\right]$ is non-zero, 
we can find a small open neighborhood of  $\bm{\Phi}=0$ such that $\det C\neq0$ by continuity, 
implying $C$ is invertible in this region.

Regarding Step 3, we write the Dirac bracket in the form
\[
\left\{ f,\, g\right\} _{DB}=\frac{\partial f}{\partial\bm{z}}\cdot \mathbb{J}_{D}\frac{\partial g}{\partial\bm{z}},
\]
where $\mathbb{J}_{D}$ is the Poisson matrix\cite{ChandreBrackets2012} satisfying the Jacobi
identity, 
\begin{equation}
0=\sum_{l}\mathbb{J}_D^{al}\partial_{l}\mathbb{J}_D^{bc}+\mathbb{J}_D^{bl}\partial_{l}\mathbb{J}_D^{ca}+\mathbb{J}_D^{cl}\partial_{l}\mathbb{J}_D^{ab}.\label{eq:Matrix jacobi identity}
\end{equation}
As explained above, we single out a subset of the $\bm{z}$ variables to be eliminated by the reduction procedure (denoted $\bar{\bm z}$) and label the remaining variables $\tilde{\bm z}$. We rewrite the $m$ constraints in the form
$$
\Phi_j({\bm z})=\bar{z}_j-\varphi_j(\tilde{\bm z}).
$$ 
This should be possible at least locally under the hypothesis of the implicit function theorem. Note that all the constraints considered in this manuscript are already of this form. Next we consider
the general coordinate change $\bm{z}=\left(\bar{\bm{z}},\tilde{\bm{z}}\right)\mapsto\left(\bm{\Phi},\bm{w}\right)$,
where $\bm{\Phi}$ are the constraints. This change of variables is invertible, and its inverse is given by $\bar{z}_j=\Phi_j+\varphi_j({\bm w})$ and $\tilde{\bm z}={\bm w}$. Since the constraints are Casimir invariants of the Dirac bracket, the transformed Jacobi matrix must be in the form
\begin{equation}
\tilde{\mathbb{J}}_{D}=\left(\begin{array}{cc}
0 & \cdots \\
\vdots & \tilde{\mathbb{J}}_{D}^{r}
\end{array}\right),\label{eq:Reduced DB matrix}
\end{equation}
where the number of rows and columns of zeros is the same as the number
of constraints. Since $\tilde{\mathbb{J}}_{D}$ satisfies the Jacobi
identity, it is straightforward to see from Eq.~\eqref{eq:Matrix jacobi identity}
that $\tilde{\mathbb{J}}_{D}^{r}$ must also. 
Setting $\bm{\Phi}$ to $0$ and
applying the coordinate change $\bm{w}\mapsto \tilde{\bm{z}}$ leads to a Poisson bracket
in $\tilde{\bm{z}}$ that satisfies the Jacobi identity. The set of functions depending on $\tilde{\bm z}$ constitutes a Poisson algebra with the Poisson bracket given by
$$
\{f,g\}_{r}=\frac{\partial f}{\partial\tilde{\bm{z}}}\cdot \mathbb{J}_{D}^{r}\frac{\partial g}{\partial\tilde{\bm{z}}}.
$$

This bracket is obtained by simply dropping
the $\partial/\partial\bar{\bm{z}}$ terms in the Dirac bracket. 
For consistency we require this to be equal to the operation of the reduced Dirac bracket 
$\left\{f_r,g_r\right\}_{DB}^r$, a condition which is obviously ensured if we drop $\partial/\partial\bar{\bm{z}}$ and use $\bar{\bm{z}}=\bar{\bm{z}}\left(\bm{\Phi},\bm{w}\right)$
to eliminate $\bar{\bm{z}}$.
Note that this is equivalent to removing the rows
and columns of $\mathbb{J}_{D}$ corresponding to the $\bar{\bm{z}}$
variables. Of course, one can reach the same point by the variable change 
$\bm{z}\mapsto\left(\bm{\Phi},\bm{w}\right)$
as explained above, illustrating that dropping $\bar{\bm{z}}$ partial
derivatives will lead to a bracket satisfying the Jacobi identity. Note that if a constraint
is simply a co-ordinate $\Phi_i=z_j$, all $\partial/\partial z_j$ will be 
automatically eliminated from the Dirac bracket by the standard Dirac procedure. 
This behavior is seen in both the finite dimensional example below as well as the
infinite dimensional brackets in the main body of the paper. For example, the functional derivative
$\delta/\delta\bm{M}_{v}$ cancels in the calculation leading to the Vlasov-Maxwell bracket, Eq.~\eqref{eq:Full VM bracket}.

We now give an example of the procedure
for a well known finite dimensional system, the Poisson bracket for
the Lorentz force. Using a 12-dimensional extended phase space $\left(\bm{x},\bm{v},\bm{p_{x}},\bm{p_{v}}\right)$
and the modified Dirac theory above, both the $\left(\bm{x},\bm{v}\right)$
bracket and the canonical bracket are easily derived in one step.
Start with the Lagrangian for a particle in an electromagnetic field (using unit charge and mass for clarity),
\begin{equation}
L=\left(\bm{A}+\bm{v}\right)\cdot\dot{\bm{x}}-\frac{1}{2}v^{2}-\phi\label{eq:SP lagrangian}
\end{equation}
and apply the Dirac theory of constraints to the canonical Poisson bracket in extended phase space
$$
\{f,g\} =  \frac{\partial f}{\partial x_{i}}\frac{\partial g}{\partial p_{xi}}-v.v. +\frac{\partial f}{\partial v_{i}}\frac{\partial g}{\partial p_{vi}}-v.v.,
$$
where $v.v.$ means switch $f$ to $g$ and vice versa. 
There are 6 constraints on the canonical momenta, $\Phi_{xi}=p_{xi}-\left(A_{i}+v_{i}\right)=0$,
$\Phi_{vi}=p_{vi}=0$ obtained from $\partial L/\partial \dot{z}_{i}=p_{zi}$.
This gives the constraint matrix $C=\mathrm{d}\left(A+v\right)$, the symplectic
form in $\left(x,v\right)$ space. The Dirac bracket follows from the computation of $C^{-1}$~:
\begin{align}
\left\{ f,g\right\} _{DB} & =\frac{\partial f}{\partial x_{i}}\frac{\partial g}{\partial p_{xi}}-v.v.+\frac{\partial f}{\partial x_{i}}\frac{\partial g}{\partial v_{i}}-v.v.\nonumber \\
+ & \frac{\partial f}{\partial p_{xj}}\frac{\partial g}{\partial v_{i}}\frac{\partial A_{i}}{\partial x_{j}}-v.v.+\bm{B}\cdot\frac{\partial f}{\partial\bm{v}}\times\frac{\partial g}{\partial\bm{v}}.\label{eq:Full FD DB}
\end{align}
Note that,
as explained above, the terms involving $\partial/\partial p_{vi}$
cancel automatically. We can now form the reduced Dirac bracket in
the variables of our choice (except $\bm{p}_{v}$) by simply removing
partial derivatives and substituting in constraints. Here there is no need to apply the constraints on $C$ before its inversion, so Step 1 has been skipped. For instance,
removing $\bm{p_{x}}$ trivially leads to the standard $\left(\bm{x},\bm{v}\right)$
bracket 
\begin{equation}
\left\{ f\left(x,v\right),g\left(x,v\right)\right\} _{DB}=\frac{\partial f}{\partial x_{i}}\frac{\partial g}{\partial v_{i}}-v.v.+\bm{B}\cdot\frac{\partial f}{\partial\bm{v}}\times\frac{\partial g}{\partial\bm{v}},\label{eq:xv bracket SP}
\end{equation}
while removing $\bm{v}$ leads to the  $\left(\bm{x},\bm{p_{x}}\right)$ canonical
bracket, 
\begin{equation}
\left\{ f\left(x,p_{x}\right),g\left(x,p_{x}\right)\right\} _{DB}=\frac{\partial f}{\partial x_{i}}\frac{\partial g}{\partial p_{xi}}-v.v.\label{eq:xpx bracket SP}
\end{equation}
One could even form a $\left(\bm{v},\bm{p_{x}}\right)$ bracket
if desired, although the $\partial A_{i}/\partial x_{j}$ term would
have to be changed into $\left(\bm{v},\bm{p_{x}}\right)$ co-ordinates
using the constraint equations. 

While these finite dimensional brackets can be straightforwardly derived
from the Lagrangian {[}Eq.~\eqref{eq:SP lagrangian}{]} through other
means, derivation of infinite dimensional brackets can be more challenging.
So long as suitable care is taken, the method outlined above can
be very useful in deriving Poisson brackets from certain types of
field theoretic Lagrangians, as is carried out in the main body of this work for the Vlasov-Maxwell 
and gyrokinetic brackets.

\section{Jacobi identity for the gyrokinetic Poisson bracket \label{Appendix JI Proof}}

We provide a direct proof of the Jacobi identity,
\begin{equation}
\left\{ \Gamma,\left\{ \Theta,\Lambda\right\} \right\} +\circlearrowright=0,\label{Jac Id, Appendix}
\end{equation}
for the gyrokinetic bracket presented above,
\begin{equation}
\left\{ \Gamma,\Theta\right\} =\int d\bm{z}\hat{F}\left\{ \frac{\delta\Gamma}{\delta\hat{F}},\frac{\delta\Theta}{\delta\hat{F}}\right\} _{sp}.\label{eq:GK bracket appendix}
\end{equation}
Here $\circlearrowright$ denotes the permutation of $\Gamma$, $\Theta$,
and $\Lambda$ through the other two possibilities. The bracket, Eq.~\eqref{eq:GK bracket appendix},
is in the form 
\begin{equation}
\left\{ \Gamma,\Theta\right\} =\left\langle \left.\frac{\delta\Gamma}{\delta\hat{F}}\right|\mathscr{Q}\:\frac{\delta\Theta}{\delta\hat{F}}\right\rangle,\label{eq:General bracket form Appendix}
\end{equation}
where $\mathscr{Q}$ is an anti-self adjoint operator with dependence
on $\hat{F}$. For operators of this type (see Ref.~\onlinecite{1982AIPC...88...13M})
the second functional derivatives cancel in the calculation of Eq.~\eqref{Jac Id, Appendix}.
This implies that the only part of $\delta\left\{ \Theta,\Lambda\right\} /\delta\hat{F}$
that contributes in the Jacobi identity is $\left\{ \delta\Gamma/\delta\hat{F},\delta\Theta/\delta\hat{F}\right\} _{sp}$
and Eq.~\eqref{Jac Id, Appendix} becomes
\begin{equation}
\int d\bm{z}\hat{F}\left\{ \frac{\delta\Gamma}{\delta\hat{F}},\left\{ \frac{\delta\Theta}{\delta\hat{F}},\frac{\delta\Lambda}{\delta\hat{F}}\right\} _{sp}\right\} _{sp}+\circlearrowright=0.\label{eq: Jac Id explicit appendix}
\end{equation}
Thus, the Jacobi identity of the functional Poisson bracket follows
directly from that for the single particle bracket.

%

\end{document}